\documentclass[twocolumn,tighten]{aastex63}

\usepackage{amsmath}
\usepackage{breakurl}

\shorttitle{Collisional Ionization in 4U~1626$-$67}
\shortauthors{Schulz et al.}

\begin{document}

\title{Collisional Ionization in the X-ray Spectrum of the Ultracompact Binary 4U~1626$-$67}

\author{Norbert S. Schulz}
\affiliation{MIT Kavli Institute for Astrophysics and Space Research, 
Massachusetts Institute of Technology, Cambridge, MA 02139, USA}

\author[0000-0001-8804-8946]{Deepto Chakrabarty}
\altaffiliation{Visiting Professor, Institute for Theory and Computation,\\ 
Harvard-Smithsonian Center for Astrophysics, Cambridge, \\MA 02138, USA.}
\affiliation{MIT Kavli Institute for Astrophysics and Space Research, 
Massachusetts Institute of Technology, Cambridge, MA 02139, USA}

\author[0000-0002-6492-1293]{Herman L. Marshall}
\affiliation{MIT Kavli Institute for Astrophysics and Space Research, 
Massachusetts Institute of Technology, Cambridge, MA 02139, USA}


\begin{abstract}
We report on high-resolution X-ray spectroscopy of the ultracompact X-ray binary pulsar
4U 1626$-$67 with {\em Chandra}/HETGS acquired in 2010, two years after the pulsar
experienced a torque reversal. The well-known strong Ne and O emission lines with
Keplerian profiles are shown to arise at the inner edge of the magnetically-truncated 
accretion disk. We exclude a photoionization model for these lines based on the absence
of sharp radiative recombination continua. Instead, we show that the lines 
arise from a collisional plasma in the inner-disk atmosphere, with $T\simeq 10^7$~K and 
$n_e\sim 10^{17}$~cm$^{-3}$. We suggest that the lines are powered by X-ray heating of the
optically-thick disk inner edge at normal incidence. Comparison of the line profiles in 
HETGS observations from 2000, 2003, and 2010 show that the inner disk radius decreased by 
a factor of two after the pulsar went from spin-down to spin-up, as predicted by magnetic 
accretion torque theory. The inner disk is well inside the corotaton radius during
spin-up, and slightly beyond the corotation radius during spin-down.  Based on the disk radius and accretion torque measured during steady spin-up, the pulsar's X-ray luminosity is $(2.0^{+0.2}_{-0.4})\times 10^{36}$~erg~s$^{-1}$, yielding a source distance of 
$3.5^{+0.2}_{-0.3}$~kpc. The mass accretion rate is an order of magnitude larger than expected from gravitational radiation
reaction, possibly due to X-ray heating of the donor. The line 
profiles also indicate a binary inclination of 39$^{+20}_{-10}$~degrees, consistent with a 
$\simeq$0.02~$M_\odot$ donor star. Our emission measure analysis favors a He white dwarf or a highly-evolved H-poor main sequence remnant for the donor star, rather than a C-O or O-Ne white dwarf. The measured Ne/O ratio is 0.46$\pm$0.14 by number. In an appendix, we show how to express the emission measure of a hydrogen-depleted collisional plasma without reference to a hydrogen number density.
\end{abstract}


\section{INTRODUCTION}

The accretion-powered pulsar 4U~1626$-$67 is a rare example of a strongly
magnetized ($\sim 10^{12}$ G) neutron star in a low-mass X-ray binary
(NS/LMXB), a striking comparison to the weakly magnetized ($\sim 10^8$ G) neutron
stars that are usually found in LMXBs.  The bright X-ray source was first
identified by {\em Uhuru} \citep{giacconi1972} and was soon shown to be an
X-ray pulsar with a period of 7.68 s~\citep{rappaport1977}.  Since its
discovery, its X-ray behavior and spin evolution have been monitored
in both the soft and hard X-ray bands by many missions
\citep[e.g.,][]{pravdo1979,nagase1984,kii1986,levine1988,shinoda1990,
mavromatakis1994,angelini1995,chakrabarty1997,camero2010}.  The pulsar's
surface dipole magnetic field strength of $3\times 10^{12}$~G is
directly measured via an X-ray cyclotron line \citep{orlandini1998,
  coburn2002, iwakiri2012, dai2017, iwakiri2019}.  

The presence of an accretion disk in 4U~1626$-$67 is well established.  The
optical counterpart KZ TrA shows pulsations at the X-ray period,
interpreted as X-ray reprocessing in the disk
\citep{mcclintock1977,grindlay1978,ilovaisky1978,mcclintock1980}.
Double-peaked emission lines in the X-ray \citep{schulz2001} and the
ultraviolet \citep{homer2002} are indicative of Keplerian disk
motion. Quasi-periodic oscillations (QPOs) in both X-rays
\citep{shinoda1990, owens1997, kommers1998,kaur2008} and the optical band
\citep{chakrabarty1998,chakrabarty2001,raman2016} are also thought
to arise in the accretion disk. Finally, the long-term spin evolution
of the pulsar is indicative of magnetic disk accretion
\citep{levine1988,chakrabarty1997,bildsten1997,camero2010,takagi2016}. 

\begin{deluxetable*}{rccccll}[ht]
\tablecaption{CHANDRA/HETGS X-RAY OBSERVATIONS \label{tab:obs}}
\tablehead{{\em Chandra} & \multicolumn{2}{c}{Start date}& \colhead{Exposure} &
  \colhead{Count rate\tablenotemark{a}} & & \\ 
  \colhead{ObsID} & (TT) & (MJD) & (ks) & (ct s$^{-1}$) & Source state & Ref.
}
\startdata
   104 & 2000 Sep 16, 14:57:01 & 51803.623 & 39.5 & 2.41 & Faint, spin-down & 1,2,3 \\
  3504 & 2003 Jun 03, 02:30:01 & 52793.104 & 94.8 & 1.68 & Faint, spin-down & 2,3 \\
 11058 & 2010 Jan 14, 11:53:01 & 55210.495 & 76.9 & 6.80 & Bright, spin-up & 3 \\
\enddata
\tablerefs{(1) Schulz et al. 2001; (2) Krauss et al. 2007; (3) This work.}
\tablenotetext{a}{First spectral order only.}
\end{deluxetable*}

4U~1626$-$67 is also well established as an ultracompact binary (i.e., with
orbital period below 80 minutes), the only one known to contain a
strongly magnetized neutron star.  There are stringent upper limits on 
Doppler motion of the pulsar from X-ray timing measurements
\citep{levine1988,shinoda1990,chakrabarty1997,jain2007}.  A persistent
lower sideband to the pulsation peak in the optical power spectrum is
thought to arise from reprocessing in a binary companion with a
42-minute orbital period \citep{middleditch1981, chakrabarty1998,
  chakrabarty2001, raman2016}. Combined with the X-ray timing limits,
this indicates that 4U~1626$-$67 is an ultracompact binary with an extremely
low-mass companion \citep{levine1988, verbunt1990,chakrabarty1998}. 

Hydrogen-rich Roche-lobe--filling binaries have a minimum orbital
period around 80~min \citep{paczynski1981,rappaport1982}.
Ultracompact binaries must therefore have H-depleted mass donors
\citep{nelson1986, pylyser1988, pylyser1989, nelemans2010}. Indeed, there is a complete 
absence of H or He lines in the optical and ultraviolet
spectra of 4U~1626$-$67 \citep{werner2006, nelemans2006}. However, X-ray
spectroscopy reveals a strong complex of Ne and O emission lines
around 1~keV \citep{angelini1995, owens1997, schulz2001,
  krauss2007}, and ultraviolet spectroscopy indicates weak emission
lines of C and O \citep{homer2002}. Based on these measurements, the
donor may be a C-O white dwarf, or possibly an O-Ne white
dwarf \citep{schulz2001, werner2006, nelemans2006}. However, an analysis 
of binary evolution and disk stability issues favors a He white dwarf donor \citep{heinke2013}.  

The spin history of the 4U~1626$-$67 pulsar is remarkable. For many years
after its discovery in 1977, the pulsar was spun up at a nearly constant
rate by accretion until undergoing an abrupt (but unobserved) torque reversal in
1990, followed by spin-down at nearly the same rate
\citep{chakrabarty1997}. This steady spin-down continued until another
torque reversal in 2008, which has been followed by a resumption
steady spin-up \citep{camero2010, jain2010}. The X-ray flux is higher
during spin-up than during spin-down \citep{chakrabarty1997,
  camero2010}.  The overall X-ray spectral continuum shape also
correlates with the torque state \citep{camero2012}.  The complex
X-ray pulse shape is also strongly dependent upon the torque state in
a systematic way \citep{beri2014}.  These correlations suggest that
the torque reversals are accompanied by discrete, systematic changes
in the inner accretion flow properties.

In this paper, we present the first analysis of a deep,
high-resolution X-ray spectrum of 4U~1626$-$67 during the bright spin-up
state, acquired in 2010 using the High-Energy Transmission Grating Spectrometer
\citep[HETGS;][]{canizares2005} on the {\em Chandra X-ray Observatory}.  We also reanalyzed
two previous deep HETGS spectra taken in 2000 and 2003 during the faint spin-down state
for comparison. The 2000 and 2003 observations were previously analyzed by 
\citet{schulz2001} and \citet{krauss2007}.  Preliminary reports on some of our 
results for the 2010 observation have been presented elsewhere \citep{schulz2011,schulz2013}.

\section{OBSERVATIONS AND ANALYSIS}

A summary of the three {\em Chandra}/HETGS observations of 4U 1626$-$67 we
analyzed is given in Table~\ref{tab:obs}. The data were reduced using the CIAO
X-ray data analysis package, along with the most recent calibration
(CALDB) products and processing procedures from the {\em Chandra}
Transmission Grating Catalog and Archive\footnote{See
  \url{http://tgcat.mit.edu}} \citep[TGCat;][]{huenemoerder2011}.
X-ray spectral analysis was performed using the ISIS\footnote{See
  \url{http://space.mit.edu/ASC/ISIS}} package, along with spectral
model functions imported from XSPEC. Uncertainties were 90\%
confidence limits calculated using  the multi-parameter grid search
utility {\tt conf$\_$loop} in ISIS. 

The zeroth-order point-spread function (PSF) of the 2010 observation
was mildly affected by photon pileup. An improved zeroth-order
position was determined using the ISIS tool {\tt findzo.sl}, which uses the
intersection of the PSF read-out streak and the HETG dispersion tracks.

\bigskip

\bigskip
\subsection{Light Curve and X-Ray Flux 
\label{cfluxes}}

\begin{figure}
\includegraphics[angle=0,width=0.47\textwidth]{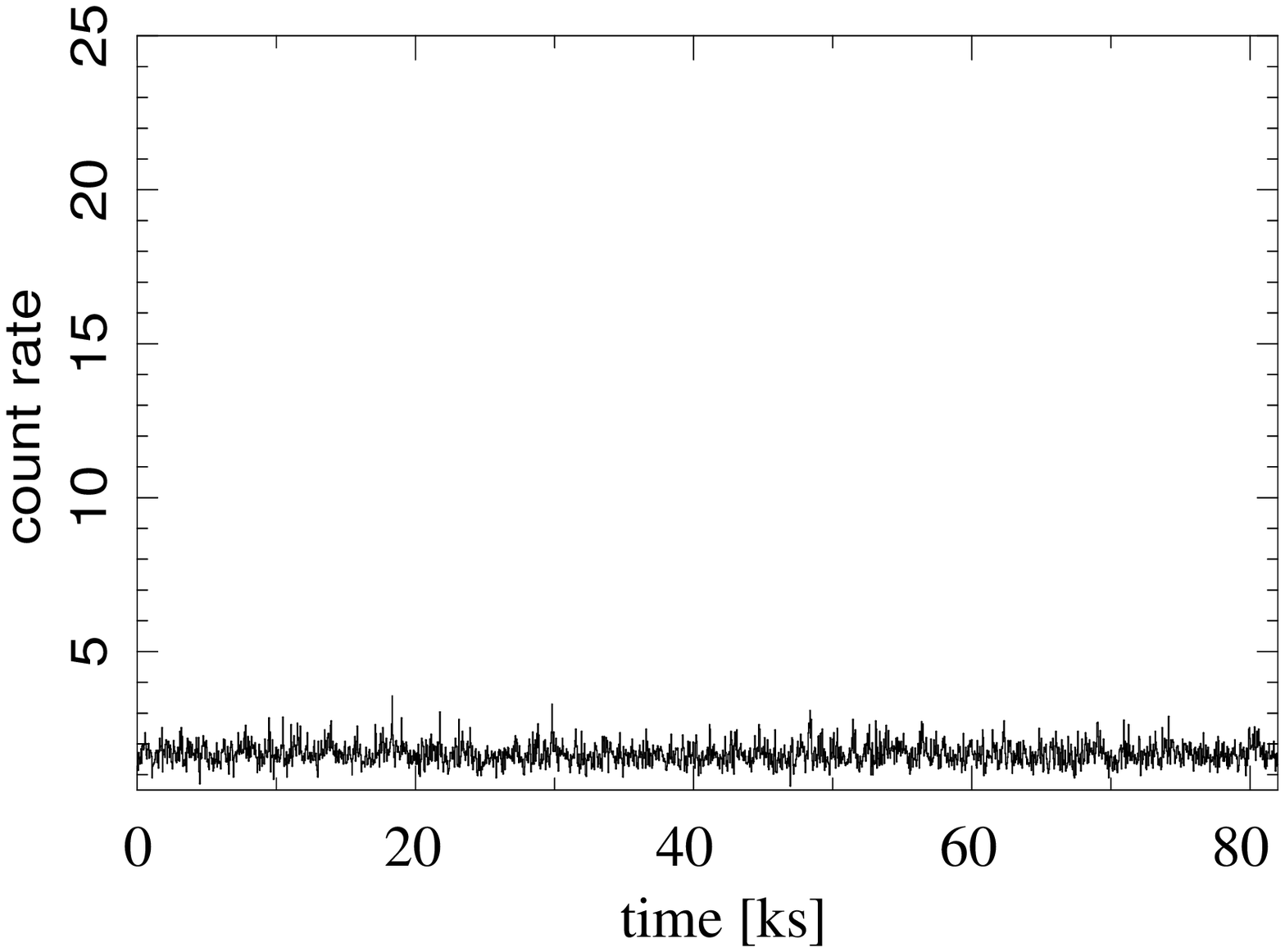}
\includegraphics[angle=0,width=0.47\textwidth]{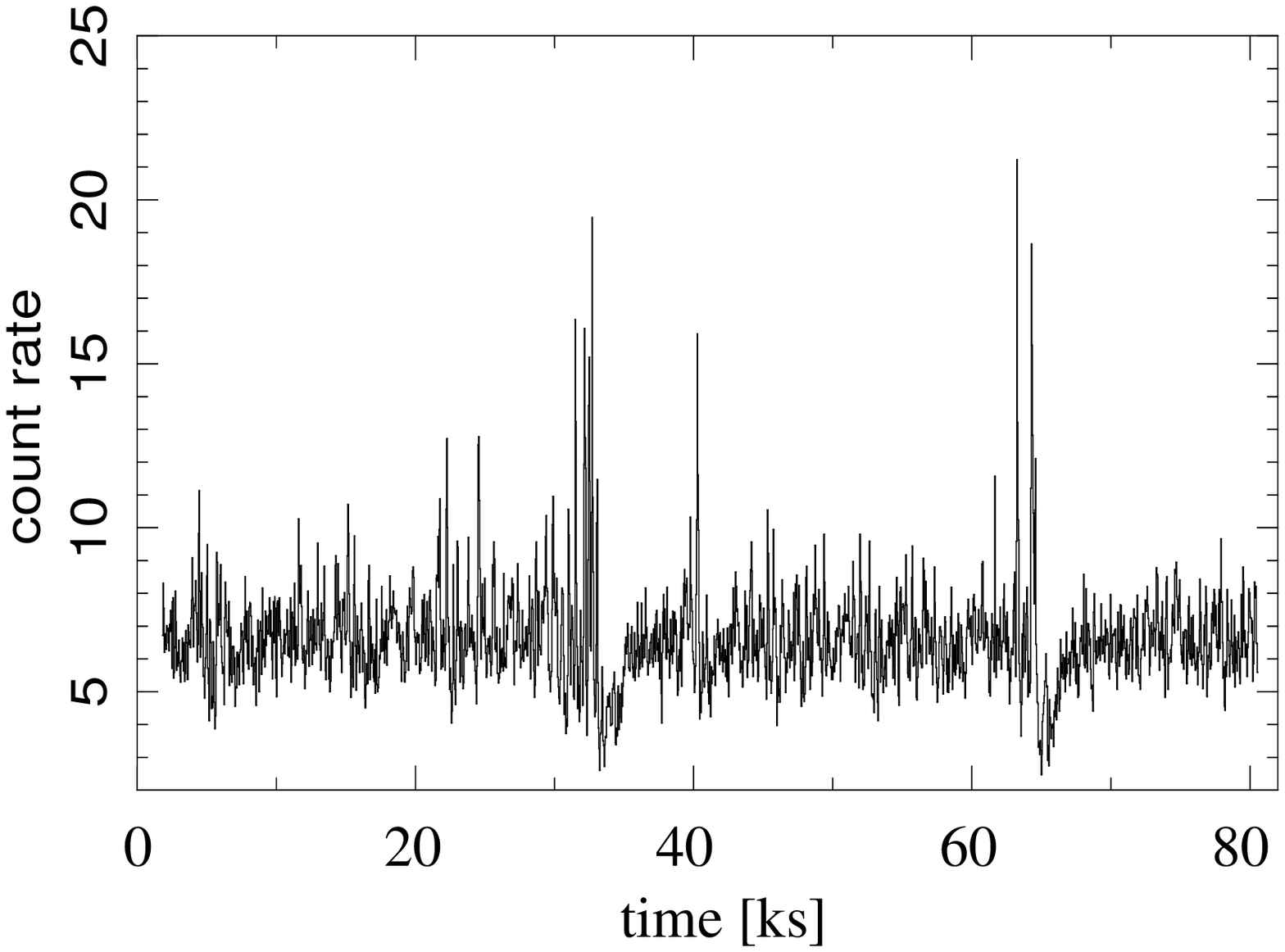}
\caption{The 0.5--10~keV X-ray light curves of 4U~1626$-$67 (from first-order
  HETGS spectra) from the two different source states, plotted on the
  same scale, binned at 40~s resolution. The count rate is plotted in
  ct~s$^{-1}$. {\bf Top:} The 2003 observation during the faint
  spin-down state. The X-ray flux is steady. {\bf Bottom:} The 2010
  observation during the bright spin-up state. There is considerable
  structure in the 2010 light curve. 
\label{fig:light1}}
\end{figure}

\begin{figure}
\includegraphics[angle=0,width=0.43\textwidth]{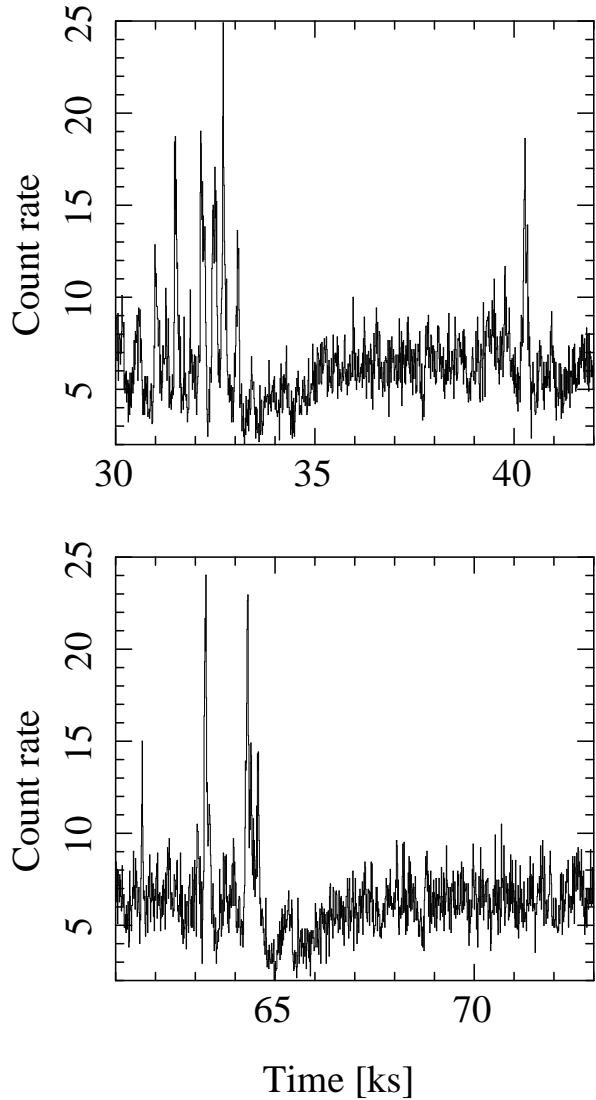}
\caption{Detailed view of the 0.5--10~keV X-ray light curve of the 2010
    observation of 4U~1626$-$67 during its bright spin-up state,
    binned at 10~s resolution.  The count rate is plotted in ct~s$^{-1}$.
    This flaring and dipping behavior is absent in observations during the 
    faint spin-down state. 
\label{fig:light2}}
\end{figure}

Figure~\ref{fig:light1} compares the 0.5--10~keV light curves of two
{\em Chandra}/HETGS observations of 4U~1626$-$67 in 10-s bins. The 2003
observation (top panel), taken during faint spin-down state, had a low
average count rate of 2.41\,ct\,s$^{-1}$ and exhibited no significant 
variability above the Poisson level. The 2010 observation (bottom
panel), taken during the bright spin-up state, had a much higher
average count rate of 6.80\,ct\,s$^{-1}$. The 2010 light curve also
contained significant structure.  There was flaring throughout the
observation, with peak intensities varying between 12 and
25~ct~s$^{-1}$. There are also two significant events at 33.5~ks and 
65~ks into the observation, which each consisted of a period of strong
flaring followed by a broad intensity dip and finally a rapid recovery
to the persistent flux level. The two events do not appear to be
associated with a strict periodicity, as they were separated by 31.5
ks and no similar event was observed at the beginning of the observation.
Figure~\ref{fig:light2} shows a detailed view of the flare/dip events from
the 2010 observation. Qualitatively, the behavior may be indicative of
a quasi-periodic cycle as follows: an interval of intermittent
weak flares, followed by successively more frequent and stronger
flares, followed by an abrupt ($<10^3$~s) factor of two dip in the persistent
flux. The dip then recovered in 1500--2000~s and was followed by another
interval of intermittent weak flaring. 

\begin{deluxetable*}{lcccc}[t!]
\tablecaption{LINE-FREE CONTINUUM SPECTRAL FITS \label{tab:contspec}}
\tablehead{ & & \multicolumn{3}{c}{Observation year} \\
\colhead{Parameter}  & \colhead{Units} & \colhead{2000} & 
    \colhead{2003} &  \colhead{2010}
}
\startdata
Absorption column density, $N_{\rm H}$ & $10^{21}$~cm$^{-2}$ & 
    1.30(14) & 1.21(15) & 1.25(5) \\
Power-law normalization at 1~keV, $A_{\rm pl}$ & 
    $10^{-2}$ ph~cm$^{-2}$~s$^{-1}$~keV$^{-1}$ & 
    1.21(1)& 0.82(1) & 3.82(2) \\
Power-law photon index\tablenotemark{a}, $\Gamma$ & \nodata & 
    0.87(1) & 0.79(1) & 1.18(1) \\
Blackbody temperature, $kT$ & keV & 0.23(1) & 0.21(1) & 0.48(1) \\
Blackbody normalization, $(R_{\rm km}/D_{\rm 10kpc})^2$ & \nodata  & 
    $405^{+82}_{-70}$ & $465^{+83}_{-72}$ & 90(4) \\
Absorbed 0.5--10~keV flux, $F$ & $10^{-10}$ erg~cm$^{-2}$~s$^{-1}$ & 
    2.2 & 1.7 & 4.6 \\
Fit statistic, $\chi^2_\nu$/dof & \nodata & 
    0.94 & 1.05 & 1.25 \\
\enddata
\tablenotetext{a}{$dN/dE \propto E^{-\Gamma}$.}
\end{deluxetable*}

The 2010 dataset, because of the significantly higher flux, is much
more suitable to search for pulsed emission at the 7.7~s pulse period.  We find a pulsed
fraction in the 3--4 keV band of 11.4$\pm 0.8 \%$ and lower fractions
at lower energies, consistent with average fractions reported
by~\citet{levine1988}, With the HETGS data, however, we can also search for
pulsed line emission.  For the full \ion{Ne}{10} and \ion{O}{8} lines we find
3$\sigma$ upper limits of $<6\%$ on the pulsed fraction. However,
focusing just on the Doppler peaks (i.e., regions lying $\pm$ 3000 km
s$^{-1}$ from the line centroids), the corresponding 3$\sigma$ upper
limits are $<9.8\%$ and $<11.7\%$, respectively. 

\begin{figure}[ht]
\includegraphics[angle=0,width=0.47\textwidth]{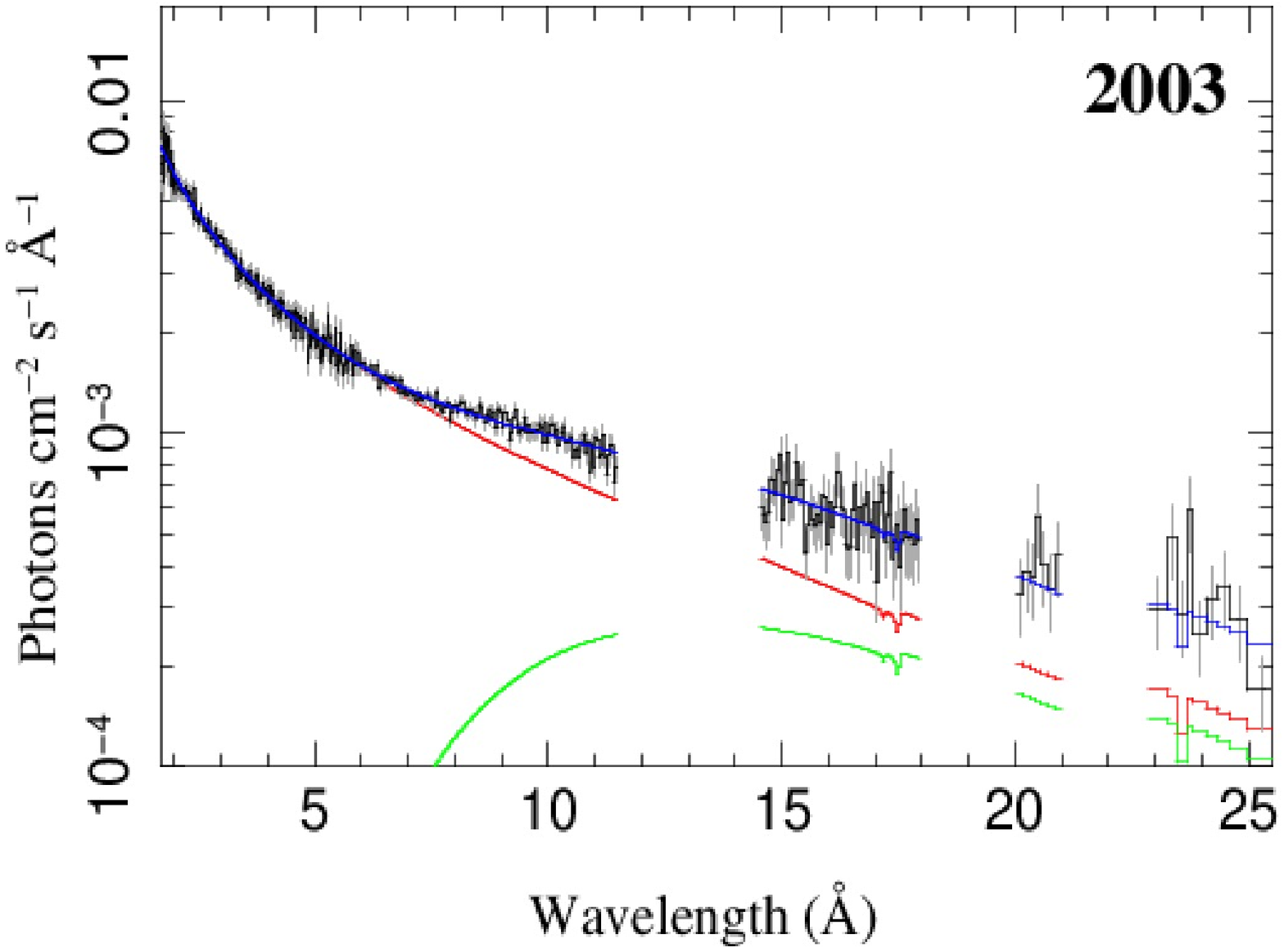}
\includegraphics[angle=0,width=0.47\textwidth]{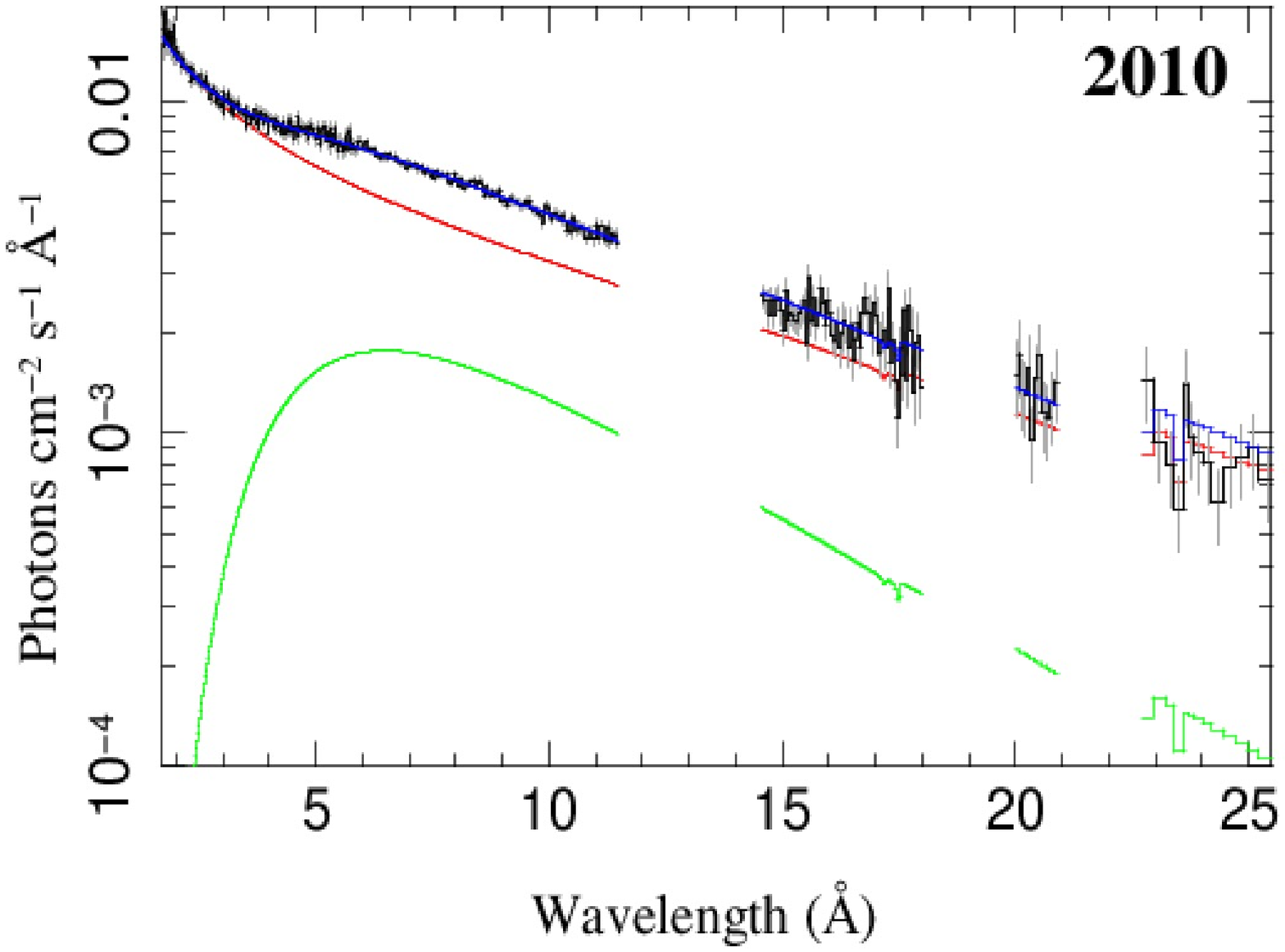}
\caption{Fits of the line-free X-ray continuum of 4U 1626$-$67. The top panel shows
  the 2003 observation during the faint spin-down state. The bottom
  panel shows the 2010 observation during the bright spin-up state. 
  The green curves are the absorbed blackbody component, and the red curves
  the absorbed power-law component.  The total absorbed model spectrum
  is shown in blue.
\label{fig:contspec}}
\end{figure}

\subsection{Continuum Spectra \label{cspectra}}

The X-ray spectrum of 4U~1626$-$67 consists of a continuum spectrum with
strong, broad line emission superimposed.  Following the procedure of 
\citet{krauss2007}, we fit the continuum spectrum by first excluding
the emission line regions.  As in most previous studies of the source,
we modeled the continuum as consisting of power-law and blackbody
components, both subject to interstellar absorption.  We used the
XSPEC {\tt tbabs} model for interstellar absorption, with the
interstellar abundance distribution of \citet{wilms2000} and the
photoelectric cross-sections of \citet{verner1996}. The
corresponding XSPEC continuum model function was {\tt
  tbabs*(bbodyrad+powerlaw)}. 

Our spectral fit parameters to the line-free continuum for all three
observations are shown in Table~\ref{tab:contspec}, and a plot of
the 2003 and 2010 observations is shown in Figure~\ref{fig:contspec}. 
Our continuum model provides a good fit to all three data sets.  The
interstellar absorption column density measured in the 2010 data is
consistent with the earlier measurements, and there were no
significant absorption edges detected in excess interstellar
absorption. However, the model parameters for the power-law and
blackbody components were significantly changed in the bright spin-up
observation of 2010 relative to the faint spin-down observations of 2000
and 2003.  The power-law slope was significantly steeper in the 2010
observation, while the blackbody temperature doubled and the
blackbody normalization was significantly smaller.  It is interesting to
note that the power-law index measured during the earlier bright spin-up
state in 1977--1990 was also steeper than typically observed during
the faint spin-down state \citep[see, e.g., Figure 3
  of][]{pravdo1979}. This supports the idea that there are distinct
X-ray spectral states associated with the pulsar accretion torque
states.


\subsection{Emission lines\label{linespec}}

A list of the spectral features of interest for 4U~1626$-$67, and whether or
not they were detected in the 2010 observation, is given in Table~\ref{tab:features}. All three
HETGS observations of 4U~1626$-$67 contain strong, broad lines of Ne and O.
However, the Ne/O emission line complex was nearly an order of magnitude
stronger in the 2010 observation than in either of the previous
observations.  This is despite the fact that the continuum flux in
2010 was only a few times brighter than previously observed.  A summary
of the Ne and O line strengths in the three HETGS observations is
given in Table~\ref{tab:lines}.  

\begin{deluxetable}{clc}
\tablecaption{2010 X-RAY SPECTRAL FEATURES \label{tab:features}}
\tablehead{ \colhead{Wavelength (\AA)}& \colhead{Feature} &
  \colhead{Detected?}
}
\startdata
  {\phn}1.94 & Fe {\sc i--x} K$\alpha$ & yes \\
  {\phn}9.15 & \ion{Ne}{10} RRC edge      & no \\
  {\phn}9.71 & \ion{Ne}{10} Ly\,$\gamma$  & no \\
       10.24 & \ion{Ne}{10} Ly\,$\beta$   & no \\
       12.13 & \ion{Ne}{10} Ly\,$\alpha$  & yes \\
       13.44 & \ion{Ne}{9} triplet ($r$)        & marginal \\
       13.55 & \ion{Ne}{9} triplet ($i$)        & yes \\
       13.70 & \ion{Ne}{9} triplet ($f$)        & marginal \\
       14.30 & \ion{O}{8} RRC edge  & no \\
       15.17 & \ion{O}{8} Ly\,$\gamma$ & no \\
       16.01 & \ion{O}{8} Ly\,$\beta$ & no \\
       18.97 & \ion{O}{8} Ly\,$\alpha$ & yes \\
       21.60 & \ion{O}{7} triplet ($r$)        & marginal \\
       21.80 & \ion{O}{7} triplet ($i$)        & yes \\
       22.10 & \ion{O}{7} triplet ($f$)        & marginal \\
\enddata
\tablecomments{The He-like triplets have three components: resonance ($r$), 
intercombination ($i$), and forbidden ($f$).}
\end{deluxetable}

\begin{deluxetable*}{llcccccccc}
\tablecaption{NEON AND OXYGEN EMISSION LINE STRENGTHS\label{tab:lines}}
\tablehead{ & & \multicolumn{4}{c}{Emission line flux\tablenotemark{a,b}} &
  \colhead{$H$} & \colhead{$G$} & \colhead{$R$} & \\ \cline{3-6}
 \colhead{Year} & \colhead{Component} & \colhead{$h$} &
 \colhead{$r$} & \colhead{$i$} & \colhead{$f$} & 
 \colhead{($h/r$)} & \colhead{($[f+i]/r$)} & \colhead{($f/i)$} & 
 \colhead{Refs.}
}
\startdata
 2000 & Ne blue wing & 15.8 & $<1.8$ & 5.0 & $<1.9$& $>9$& $>3$& $<0.4$&1 \\
      & Ne red wing  & 23.8 & $<1.8$ & 5.0 & $<1.9$&$>13$&$>3$ & $<0.4$&1 \\
      & O blue wing  & 26.3 & 12.8 & 19.8 & $<12.8$&2&$\sim 2$ & $<0.6$&1 \\
      & O red wing   & 21.6 & 12.8 & 19.8&$<12.8$&1.7&$\sim 2$&$<0.6$&1\\ \\
 2003 & Ne blue wing & 8.2 & $<2.1$ &2.3 & $<1.2$ & $>4$& $>1.1$& $<0.5$&1 \\
      & Ne red wing & 10.5 &$<2.1$& 2.3 & $<1.2$& $>5$ & $>1.1$& $<0.5$&1 \\
      & O blue wing & 13.0 & $<10$& 13.6& $<2.7$& $>1$& $>1.4$& $<0.2$&1\\
      & O red wing & 13.7 & $<10$& 13.6& $<2.7$& $>1$& $>1.4$&$<0.2$& 1\\ \\
 2010 & Ne blue wing& 83.4 & $<11.3$& 10.9& $<10.6$&$>8$& $>1$& $<1$ & 2\\
      & Ne red wing& 89.6 & $<11.3$& 10.9& $<10.6$& $>8.5$& $>1$&$<1$ & 2 \\
      & O blue wing& 97.5&$<29.0$&43.7&$<34.8$&$>3.4$& $>1.5$&$<0.8$ & 2 \\
      & O red wing& 117.0&$<29.0$&43.7&$<34.8$&$>4$&$>1.5$&$<0.8$& 2 \\
\enddata
\tablerefs{(1) Krauss et al. 2007; (2) This work.}
\tablenotetext{a}{Gaussian line flux in units of
  $10^{-5}$\,ph~cm$^{-2}$~s$^{-1}$.} 
\tablenotetext{b}{In these columns, $h$ refers to the hydrogen-like
  Ly$\alpha$ line (\ion{Ne}{10} or \ion{O}{8}); $r$, $i$, and $f$ refer to the
  helium-like triplets (\ion{Ne}{9} or \ion{O}{7}).} 
\end{deluxetable*}

The He-like \ion{Ne}{9} and \ion{O}{7} lines are triplets, and the relative
strengths of these triplet lines can provide a diagnostic of the plasma
conditions in the line forming region \citep{porquet2010}.
\citet{schulz2001} and \citet{krauss2007} previously used the He-like
triplets to investigate both the temperature and the density of the
line forming regions in 4U~1626$-$67.  They inferred an electron temperature
$T_e\gtrsim 10^6$~K and a high electron density.  However,
\citet{krauss2007} also noted that the density diagnostic power of the
He-like triplets could be spoiled by the presence of a strong
ultraviolet continuum, although they did not attempt to evaluate that
risk quantitatively.  This concern turns out to be valid.  As we
show in Appendix~A, the ultraviolet continuum expected
from the accretion disk in 4U~1626$-$67 is more than sufficient to
distort the relative strength of the He-like triplets, rendering them
unusable as a density diagnostic in this source.

\begin{figure}
\includegraphics[angle=90,width=0.47\textwidth]{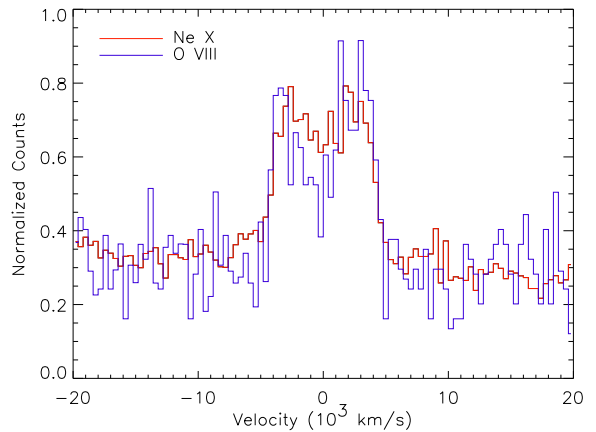}
\caption{The Keplerian line profiles of the hydrogen-like Ly$\alpha$
  lines of \ion{Ne}{10} (red) and \ion{O}{8} (blue), from the 2010
  observation. The lines are plotted on a velocity scale centered on
  their rest wavelengths.  The \ion{O}{8} flux has been re-scaled by a
  factor of 10 in order for comparison with the \ion{Ne}{10} line.
  The velocity structure of the two lines is essentially identical,
  indicating that they are both formed at approximately the same
  accretion disk radius.
\label{hlikelines}}
\end{figure}

\begin{figure}
\includegraphics[angle=0,width=0.47\textwidth]{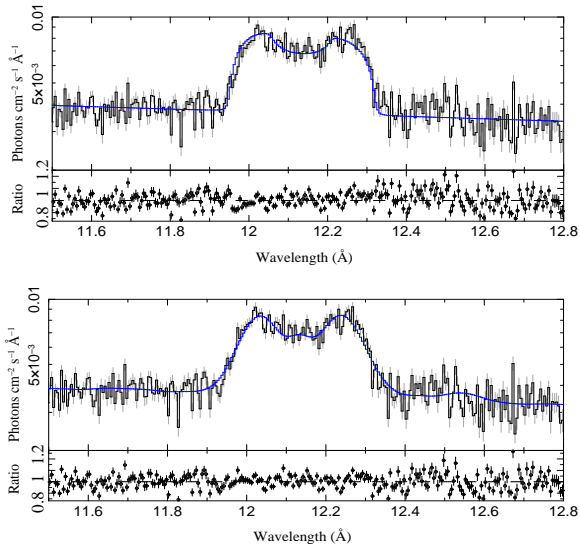}
\caption{Fits of the shape of the \ion{Ne}{10} line in the 2010
  observation. {\bf Top:} Fit using the multi-component APEC model
  described in Section~\ref{cmission}.  {\bf Bottom:} Fit using the
  {\tt diskline} model described in Section~\ref{linespec}.  Both
  models fit the data adequately.
\label{fig:diskline}}
\end{figure}

The tremendous strength of the 2010 emission lines makes them ideal
for kinematic analysis.  We confined our attention to the intrinsically
singlet hydrogen-like Ly$\alpha$ lines of \ion{Ne}{10} and
\ion{O}{8}. Figure~\ref{hlikelines} shows the profiles of these lines in
velocity space. For both line profiles, we observed a velocity range of
$\pm$4000 km s$^{-1}$. The line shapes are strongly indicative of
Doppler-shifted line pairs with a Keplerian profile. When fitted by a
pair of Gaussians, we find line shifts of $\pm$2200~km~s$^{-1}$ and a
full-width at half-maximum (FWHM) of 1800~km~s$^{-1}$, which are
consistent with previous observations \citep{krauss2007}. Both lines
have similar kinematic properties, indicating that they are both
formed at around the same radius in the accretion disk.  

\begin{deluxetable*}{lccccc}[t]
\tablecaption{DISK LINE PROFILE FITS TO HYDROGEN-LIKE LINES\label{tab:diskline}}
\tablehead{ & & \multicolumn{2}{c}{2010 (spin-up)} & 
  \multicolumn{2}{c}{Joint 2000+2003 (spin-down)} \\
  \colhead{Parameter} & \colhead{Units} & 
  \colhead{\ion{Ne}{10}} & \colhead{\ion{O}{8}} & \colhead{\ion{Ne}{10}} & \colhead{\ion{O}{8}} } 
\startdata
Central wavelength, $\lambda$ & \AA & 12.13(1) & 18.97(1) & 12.13(1) & 18.97(1) \\
Line flux, $F$ & $10^{-2}$\,ph~cm$^{-2}$~s$^{-1}$& 0.19(1) & 0.22(3) &0.017(4) & 0.018(5)\\
Emissivity index\tablenotemark{a}, $q$ &\nodata & $-$3.6 & $-$3.6 & $-$4.6& $-$4.6\\
Inner disk radius, $r_{\rm in}$ & $10^3\,GM_x/c^2$ & $1.8^{+0.2}_{-0.5}$ &
     $1.7^{+0.2}_{-0.1}$ & $3.9^{+0.7}_{-0.9}$ & $3.8^{+0.2}_{-0.6}$\\
Outer disk radius, $r_{\rm out}$ & $10^3\,GM_x/c^2$& $10.9^{+10.2}_{-2.8}$ & 
     $3.8^{+1.3}_{-0.6}$ & $94.3^{+5.6}_{-81.0}$& $14.6^{+7.4}_{-10.6}$\\
Disk inclination, $i_d$ & degrees & $39.0^{+10.2}_{-5.2}$ & (tied) &(tied) &(tied) \\
Fit statistic, $\chi^2_\nu$/dof & \nodata & 1.09 & (tied) &(tied) &(tied) \\
\enddata
\tablenotetext{a}{Line emissivity scales with disk radius as as $r^q$.}
\tablecomments{Lines are fit to the XSPEC {\tt diskline} model
  \citep{fabian1989}. Single-parameter uncertainties are quoted. However, 
  $r_{\rm in}$ and $i_d$ are highly correlated (see~Figure~\ref{fig:contour}).} 
\end{deluxetable*}

We further examined the Keplerian nature of these line shapes by
fitting them with the XSPEC {\tt diskline} model for line emission
from a Keplerian accretion disk \citep{fabian1989}. The parameters for
this model are the line wavelength, the disk inclination angle $i_d$,
the inner and outer disk radii ($r_{\rm in}$ and $r_{\rm out}$) for the
line emitting region, and the line emissivity index $q$ (where the
line emissivity scales with disk radius as $r^q$). Although the model
was originally designed for use with fluorescent lines from a black hole
accretion disk, it should be generally applicable to any disk line
emission (although possibly with  different emissivity behavior). We
fit a narrow region around each of the hydrogen-like lines in the 2010
data using the XSPEC model {\tt powerlaw+diskline}. We tied $i_d$ and 
$q$ between the two line fits, but allowed for separate disk radii. We
found that choosing $q=-3.6$ gave the best fits for the 2010 data, and
so held $q$ fixed at this value.  The model fit the data
well. Figure~\ref{fig:diskline} shows the fit to the 2010
\ion{Ne}{10} line. The parameters $\lambda_0$, $i_d$, $r_{\rm in}$ were well
determined, but the outer radius $r_{\rm out}$ was essentially
unconstrained.  This is not surprising, since the outer radius
produces the smallest Doppler shifts. 

\begin{figure}
\includegraphics[angle=0,width=0.47\textwidth]{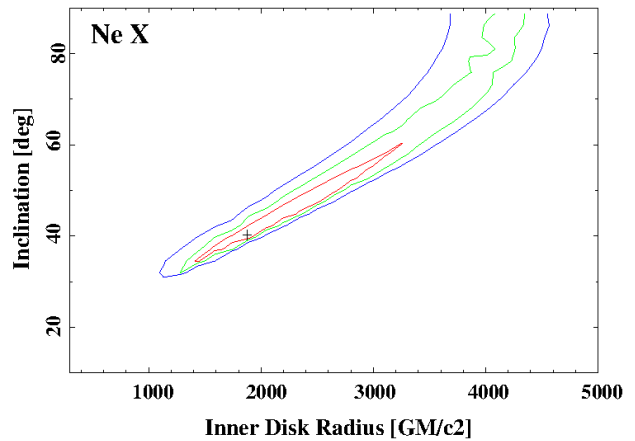}
\caption{Contour plot for the {\tt diskline} fit for the \ion{Ne}{10} line
from the 2010 observation, showing disk
  inclination angle $i_d$ versus inner disk radius $r_{\rm in}$. The red, green, blue
  contours represent 1, 2, and 3 sigma contours, respectively. Accounting for the
  correlation of $i_d$ and $r_{\rm out}$, the best-fit disk inclination is
  $i_d=39^{+20}_{-10}$~degrees.
\label{fig:contour}}
\end{figure}

We also attempted the same analysis on the earlier data from the faint
spin-down state.  We combined the 2000 and 2003 line spectra in order
to improve the statistics.  We fit these combined data simultaneously
with the 2010 data, requiring the same inclination angle for all the data but
allowing for different disk radii for each line and the two spin-state epochs
(2000+2003 and 2010).  We held $q$ fixed at $-3.6$ for the 2010 
data and $-4.6$ for the 2000+2003 data, based on preliminary fitting
results. The combined best-fit parameters are given listed in
Table~\ref{tab:diskline}. The inner disk radius of the emitting region
is consistent between the Ne and O lines for each observation, with the
radius decreasing by a factor of two between spin-down (2000+2003) and
spin-up (2010).   The fits for $i_d$ and $r_{\rm in}$ are highly correlated, 
as shown in Figure~\ref{fig:contour}. Accounting for this correlation, the 
best-fit disk inclination is $i_d= 39^{+20}_{-10}$~degrees (or equivalently
$\sin i = 0.63^{+0.23}_{-0.15}$).

\begin{figure}
\includegraphics[angle=0,width=0.47\textwidth]{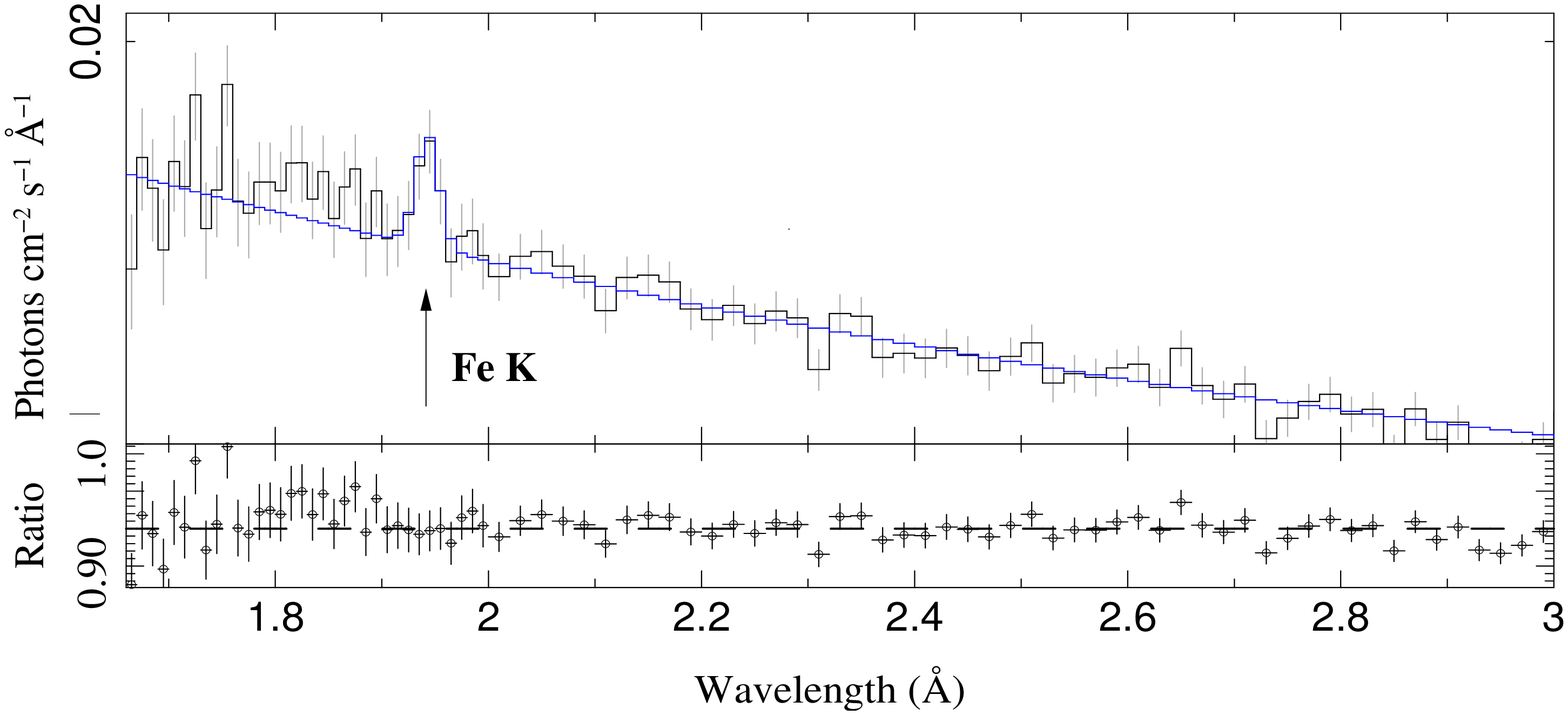}
\caption{The fluorescent Fe K line in the 2010 observation.
\label{fekline}}
\end{figure}

We detected a weak emission line feature in the Fe K line region of
the 2010 observation.  The Fe~line detection in these data has already 
been reported independently by
\citet{koliopanos2016}. Figure~\ref{fekline} shows the continuum fit in the
1.5--3~\AA\ region.  A Gaussian line fit gives a wavelength of
1.94$\pm$0.06~\AA~(6.4 keV) and an equivalent width of about
6~m\AA~(20 eV). The line appeared unresolved, with an upper limit on
the line width of $\sigma$ = 9~m\AA.  The line location was consistent
with fluorescence from near neutral ionization states,
i.e. \ion{Fe}{1}--\ion{Fe}{10}. The fit leaves a broad residual at shorter
wavelengths consistent with \ion{Fe}{25}, but its significance is less
than 2 $\sigma$. The fluorescence line appeared to be persistent.  It
was not detected separately during the flares, but this is probably due to the low
signal-to-noise ratio during these events. There were no other 
detections of line fluorescence in the spectrum. The line was not
detected in previous {\em Chandra} observations. Our detection in the
2010 observation is consistent with other reports of weak Fe line
emission in 4U~1626$-$67 since the 2008 torque reversal and X-ray brightening
\citep{camero2012, koliopanos2016,dai2017,iwakiri2019}. 

\section{Ionization Modeling\label{ionmod}}

Previous HETGS studies of 4U~1626$-$67 modeled the emission lines as
double-Gaussian lines \citep{schulz2001,krauss2007}, without
attempting to account for their relative strengths by modeling the
underlying plasma conditions.  Many steady-state astrophysical plasmas
are in either photoionization equilibrium or collisional ionization
equilibrium, although a hybrid state is possible as well \citep{pradhan2011}.  
However, the accretion disk atmosphere and corona of a
NS/LMXB like 4U~1626$-$67 is usually assumed to be photoionized, since the disk is known
to be illuminated by a strong central X-ray continuum source
\citep{jimenez2002}.  In this section, we explicitly examine both scenarios
for 4U~1626$-$67. 

\subsection{Photoionized Emission\label{pmission}}

In a photoionized plasma, the physical conditions are set by the
incident radiation field \citep[see,
  e.g.,][]{mewe1999,liedahl1999,pradhan2011}.  This is described by
the ionization parameter $\xi = L/n_e\,r^2$, where $L$ is the ionizing
luminosity from a source at distance $r$, and $n_e$ is the electron
density.  We modeled the line-emitting plasma in the 2010 observation
in ISIS using the XSPEC model {\tt photemis}, a variant of the {\tt warmabs} 
model provided by the XSTAR 
code\footnote{See \url{https://heasarc.nasa.gov/docs/software/xstar/xstar.html}} 
for modeling of photoionized plasmas \citep{kallman2001}. The standard version
of this model assumes that the ionizing radiation has a $\Gamma=2$ power-law 
spectrum; this is a slightly steeper power-law than what is actually observed, 
but it is close enough for our purposes.  

The double-peaked Keplerian lines were modeled as pairs (blue and red wings) 
with equal and opposite Doppler shifts.  We found that a
single ionization parameter value could account for either the hydrogen-like
lines of O and Ne or the helium-like lines, but not both the hydrogen-like
and helium-like features at the same time.
Instead, we used a four-component photoionization model, consisting of 
two ionization parameters (for H-like and He-like features), with each 
having two line wings (blue and red).  These components were combined 
with the continuum model described earlier to
produce a global model, equivalent to the XSPEC model 
{\tt tbabs*(bbodyrad + powerlaw  + photoemis(1) + photoemis(2) + photoemis(3) + photoemis(4))}.

\begin{figure}
\includegraphics[angle=0,width=0.47\textwidth]{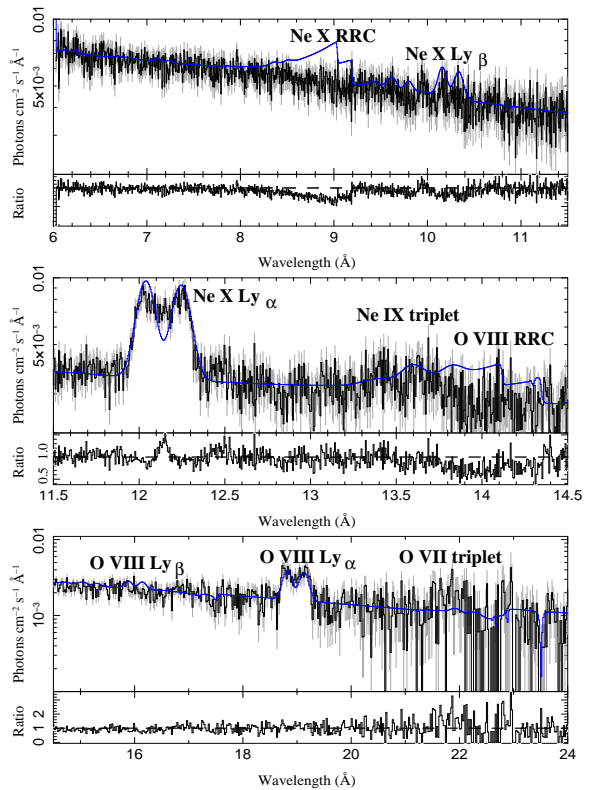}
\caption{Spectral fit of the 2010 observation using a photoionization
  model. The fit quality is poor. The predicted radiative recombination continuum edges of
  \ion{Ne}{10} and \ion{O}{8} are conspicuously absent.
\label{photspec}}
\end{figure}

The best-fit model required $\log\xi=2.2$ (cgs) for the He-like lines and
$\log\xi=2.6$ for the H-like lines.  However, the fit was poor, with a
reduced $\chi^2_\nu$ value of over 5. The residuals are shown in
Figure~\ref{photspec}. The most serious problems are the
overprediction of both the radiative recombination continuua (RRC) and
the Ly$\beta$ lines of \ion{Ne}{10} and \ion{O}{8}.  The \ion{Ne}{10} RRC residual is
especially pronounced. The suppression of RRC features corresponds to a 
high electron temperature $T_e$ (and hence a large $\log\xi$), such that $kT_e$ is
comparable to or larger than the ionization energy of the recombined state 
\citep{hatchett1976,liedahl1996}, which is 1.3~keV for \ion{Ne}{10}. 
Our fit values of $\log\xi$ correspond to a very low temperature 
$kT_e\simeq 20$~eV \citep{kallman2001},
which is why the RRC features are so sharp in the model. An ionization parameter
large enough to suppress the RRCs would not be able to reproduce the observed
strengths of the H-like lines.  This issue is unrelated to the slight mismatch in the slope of the photoionizing continuum assumed in our model.  It is simply not possible to reconcile our observed line strengths with the simultaneous absence of RRCs using a photoionization model.  We found the same problem in the 2000 and 2003 data, though
with lower significance.  We conclude that a pure photoionization model 
is ruled out for the line-emitting plasma in 4U~1626$-$67. 

\bigskip

\bigskip
\subsection{Collisionally Ionized Emission\label{cmission}}  

In a collisionally ionized plasma, conditions are controlled
by the electron temperature $T_e$ \citep[see, e.g.,
][]{mewe1999,pradhan2011}.  We can fit our data using one of the APEC
spectral models for optically-thin collisional plasmas, calculated using the Astrophysical Plasma Emission Database (APED\footnote{See
  \url{http://atomdb.org}}) available in both ISIS and XSPEC
\citep{smith2001}.  There are, however, some complications involved in our analysis.
In a purely collisional plasma, $T_e$ is essentially determined by
the shape of the bremsstrahlung continuum spectrum, with the elemental abundances set by the relative emission line strengths and the emission measure set by the overall
normalization. In our case, however, the bremsstrahlung emission is weak
(a few percent) compared to the blackbody and power-law continuum components from the
central X-ray source.  We must therefore include these illumination
components in our model.

Another complication is that the APEC models were designed to work with
H-rich plasmas, with nearly all the free electrons coming from the
ionization of H and He. In the case of an ultracompact binary like 4U~1626$-$67, 
we expect that the accretion disk consists of a H-free (and He-free) plasma, so that the
electrons must come from heavier elements.  Although the XSPEC model {\tt vvapec}
allows for zero abundance of H and He in generating the model
spectrum, the model normalization is still expressed in terms of the usual 
emission measure,
\begin{equation}
\mbox{\rm EM} = \int n_e\,n_{\rm H}\,dV ,
\end{equation}
which is defined in terms of the hydrogen number density, $n_{\rm H}$.
We show in Appendix B how the APEC normalization can be re-expressed
in terms of an equivalent definition of the emission measure that does not
refer to $n_{\rm H}$,
\begin{equation}
\mbox{\rm EM} = \int \beta n_e^2\,dV ,
\label{eq:EMbeta}
\end{equation}
where the value of the dimensionless constant $\beta$ will depend upon
the assumed composition of the plasma.  Note that the correction factor $\beta$ is 
calculated using the fit abundances for a specific observation. It is {\em not} generally 
a unique property of a particular plasma composition, except in the H-rich case.

\begin{figure}
\includegraphics[angle=0,width=0.47\textwidth]{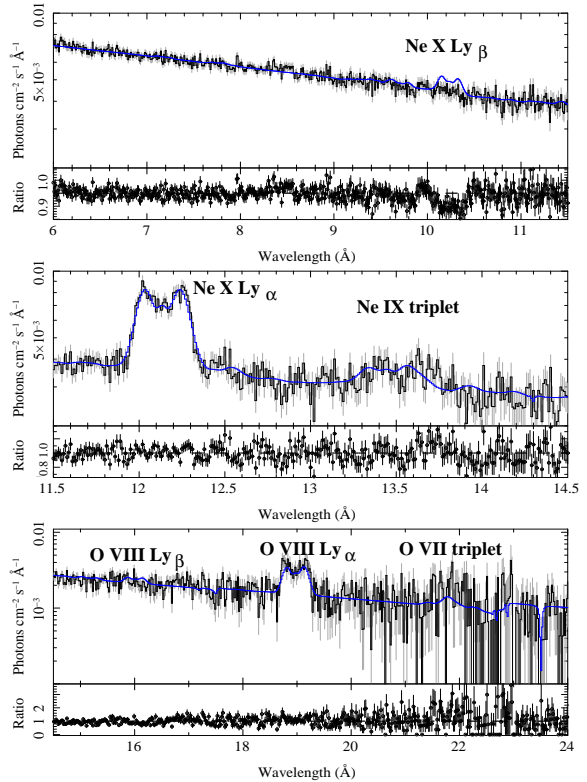}
\caption{Spectral fit of the 2010 observation using a collisional
  ionization model. The model fits very well. The \ion{Ne}{10}
  Ly$\beta$ line at 10.24~\AA\ is overpredicted. 
\label{fig:collspec}}
\end{figure}

We modeled the line emitting plasma in ISIS using the equivalent of
the XSPEC model {\tt vvapec}. The parameter Ab($Z$) is the number abundance 
of element $Z$ relative to the solar abundance model of \citet{anders1989}. An
ordinary C-O white dwarf typically contains an equal mass of C and O
\citep{segretain1994}, corresponding to $n_{\rm C}/n_{\rm O} =
16/12$. We therefore fixed Ab(C) = $3.13\times$ Ab(O) in order to achieve this
fraction. (Since there are no detectable C features in the HETGS bandpass, this
step did not end up having any practical effect.) We then set the
abundance of all elements (including H and He) to zero, {\em except}
for Ne, O, and C, and we left Ab(Ne) and Ab(O) as free parameters. 

\begin{deluxetable*}{lcccc}[t]
\tablecaption{COLLISIONAL SPECTRAL FIT PARAMETERS \label{tab:coll}}
\tablehead{ \colhead{Parameter} & \colhead{Units} & 
  \colhead{2000 Sep} & \colhead{2003 Jun} & \colhead{2010 Jan}}
\startdata
\sidehead{\underline{\em Continuum parameters}}
Interstellar column density, $N_{\rm H}$ (fixed)& $10^{22}$\,cm$^{-2}$& 
    0.12 & 0.12 & 0.12 \\
Blackbody temperature, $kT_{\rm bb}$ & keV & 0.27(1) & 0.22(1) & 0.50(1) \\
Blackbody normalization, $(R_{\rm km}/D_{\rm 10kpc})^2$ & \nodata & 
    $234^{+46}_{-37}$ & $287^{+72}_{-56}$ & 96(10) \\
Power-law photon index, $\Gamma$ & \nodata & 0.80(5) & 0.78(3) & 1.07(3) \\
Power-law normalization at 1 keV, $K_{\rm PL}$ & 
    $10^{-2}\,$ph\,cm$^{-2}$\,s$^{-1}$\,keV$^{-1}$ 
       & 1.09(6) & 0.81(3) & 3.16(12) \\
\\
\sidehead{\underline{\em APEC elemental abundances relative to solar}\tablenotemark{a}}
Hydrogen abundance, Ab(H) (fixed) & \nodata & 0  & 0  & 0  \\
Helium abundance, Ab(He) (fixed)  & \nodata & 0  & 0  & 0 \\
Carbon abundance\tablenotemark{b}, Ab(C)  (fixed)
   & \nodata & 3.13 Ab(O) & 3.13 Ab(O) & 3.13 Ab(O) \\
Oxygen abundance, Ab(O)  
   & \nodata & 1.41(36) & 1.0(2) & 0.8(2) \\
Neon abundance, Ab(Ne) 
   & \nodata & 4.6(4) & 2.58(23) & 2.6(5) \\
\\
\sidehead{\underline{\em APEC, hot components}}
Electron temperature, $T_h$ & $10^6$ K & 
     $8.5^{+2.5}_{-1.2}$ & $7.9^{+2.4}_{-0.7}$ & 10.2(8) \\
Turbulent velocity, $v_h$ & km~s$^{-1}$ & 
   $2130^{+560}_{-340}$ & $1790^{+310}_{-230}$ & 1670(120) \\
Redshift (blue and red wings), $z_h$ & $10^{-2}$ & 
   $\pm$0.60(6) & $\pm$0.58(5) & $\pm$0.87(3) \\
Normalization\tablenotemark{c}, blue wing, $K_1$ & $10^{-2}$ cm$^{-5}$ & 
   0.12(2) & 0.10(3) & 0.90(14) \\
Normalization\tablenotemark{c}, red wing, $K_2$ & $10^{-2}$ cm$^{-5}$ & 
   0.17(2) & 0.13(4) & 0.91(15) \\
Normalization\tablenotemark{c}, line center, $K_5$ & $10^{-2}$ cm$^{-5}$ & 
   \nodata & \nodata & 0.36(10) \\
\\
\sidehead{\underline{\em APEC, cool components}}
Electron temperature, $T_c$ & $10^6$ K & 
    $1.3^{+0.6}_{-0.5}$ & $1.7^{+0.3}_{-0.6}$ & $2.0^{+0.1}_{-0.3}$ \\
Turbulent velocity, $v_c$ (fixed) & km~s$^{-1}$ & 2000  & 2000  & 2000  \\
Redshift (blue and red wings), $z_c$ & $10^{-2}$ & 
   $\pm$0.9(3) & $\pm$0.55(25) & $\pm$1.22(25) \\
Normalization\tablenotemark{c}, blue wing, $K_3$ & $10^{-2}$ cm$^{-5}$ & 
   $<$0.07 & $<$0.05 & $0.02^{+0.04}_{-0.01}$ \\
Normalization\tablenotemark{c}, red wing, $K_4$ & $10^{-2}$ cm$^{-5}$ & 
   0.36(12) & 0.12(4) & $0.20^{+0.11}_{-0.07}$ \\
\\
Fit statistic, $\chi^2_\nu$/dof & \nodata & 1.01 & 1.03 & 1.21 \\
\enddata
\tablenotetext{a}{Multiplies solar abundance ratio relative to H, $(n_X/n_{\rm H})_\odot$, taken from \citet{anders1989}.}
\tablenotetext{b}{Assume equal amounts of C and O by mass, as in a C-O
dwarf.}
\tablenotetext{c}{Nominally $(10^{-14}/4\pi D^2)\int
  n_e\,n_{\rm H} dV$. See Appendix B.}
\label{tab:apec}
\end{deluxetable*}

As in the photoionized case, we modeled the double-peaked Keplerian
lines as Doppler-shifted pairs (blue and red line wings). 
A single value of $T_e$ could fit the lines of \ion{Ne}{10}, 
\ion{Ne}{9}, and \ion{O}{8}, but not \ion{O}{7} (which required a
lower temperature plasma). We therefore used a four-component
collisional model consisting of two temperatures and two line wings
(blue and red). These APEC components were combined with the continuum
model described earlier to produce a global model. For a given
temperature component, both subcomponents (blue and red) shared the
same velocity width and had equal and opposite Doppler shifts. Also,
all APEC components used the same abundances.  For the 2010 data, the
shape of the \ion{Ne}{10} line required adding a fifth APEC component,
corresponding to zero Doppler shift (for the line center).  The global
model used was thus equivalent to the XSPEC model {\tt tbabs*(bbodyrad + powerlaw +
  vvapec(1) + vvapec(2) + vvapec(3) + vvapec(4) + vvapec(5))}.

This collisional model fits the data very well. The best-fit
parameters for all three observations are listed in 
Table~\ref{tab:apec}, and the fit to the 2010 observation is shown in
Figure~\ref{fig:collspec}.  The fit correctly reproduces nearly all
the features in the spectrum. The only significant residual is a slight
overprediction of the Ly$\beta$ line of hydrogen-like \ion{Ne}{10}.  
The abundance fits imply that $n_{\rm Ne}/n_{\rm O}=0.46\pm 0.14$, 
independent of the overall plasma composition (see Appendix~B).
We conclude that the Ne and O lines in 4U~1626$-$67 are consistent with
emission from a collisionally ionized plasma.

\section{Discussion}

\subsection{Location of line-emitting region\label{sec:lineregion}}

The Keplerian profiles of the Ne and O emission lines in 4U~1626$-$67 make 
clear that the lines arise in the accretion disk. The disk in
this source extends from the pulsar magnetosphere around radius
$r\sim 10^8$~cm to the neutron star's tidal radius around
$r\approx 2\times 10^{10}$~cm \citep{chakrabarty1998}.  The similar line 
profiles observed for \ion{Ne}{10} and \ion{O}{8} show that they both arise
at the same disk radius.   

We can estimate this radius using our line profile fits for the
hydrogen-like lines of Ne and O (Table~\ref{tab:diskline}). We assume
that the lines arise in an annular region of the disk.  For the 2010
observation obtained during spin-up of the pulsar, the inner radius of
this annulus is
\begin{equation}
r_{\rm in} \approx 1800 \left(\frac{GM_x}{c^2}\right) 
   =  3.7\times 10^8\,M_{1.4} \mbox{\rm\ cm},
\end{equation} 
where we have taken $i_d=39^\circ$. 
We see that the line profile fits place the emission region at 
or near the inner edge of the accretion disk, where it is truncated by the pulsar's
magnetosphere.  

It is interesting to compare this to the radius found during spin-down of
the pulsar.  From the 2000/2003 spin-down observations, we find
\begin{equation}
r_{\rm in} \approx 4000 \left(\frac{GM_x}{c^2}\right)
   =  8.3\times 10^8\,M_{1.4} \mbox{\rm\ cm}. 
\end{equation}
There was a clear change in the location of the inner disk edge associated with the
torque reversal: $r_{\rm in}$ was more than twice as large during
spin-down than during spin-up.  This is consistent with 
standard magnetic accretion torque theory \citep[e.g.,][]{ghosh1979b},
which predicts that pulsar spin-up occurs when a large mass accretion
rate $\dot M$ (corresponding to a high X-ray luminosity) pushes the
magnetospheric boundary inward to smaller radii.  The theory further
predicts that spin-down occurs at lower $\dot M$, when the inner disk
has a larger radius, close to (but inside) the so-called corotation radius 
(where the Keplerian and pulsar rotational angular velocities are equal),
\begin{equation}
r_{\rm co} = \left(\frac{G M_x P^2}{4\pi^2}\right)^{1/3} = 
6.5\times 10^8\,M_{1.4}^{1/3} \mbox{\rm\ cm},
\end{equation}
where $P= 7.7$\,s is the pulsar spin period.  Our observations provide a clear
demonstration of this trend.  However, we formally find $r_{\rm
  in} >r_{\rm co}$ during spin down, contradicting the theoretical expectation that
$r_{\rm in} \lesssim r_{\rm co}$ for steady accretion
\citep{ghosh1979b}. In our case, this condition is only satisfied for
$M_x \lesssim 1\,M_\odot$, which is implausible for a neutron star. This
may be evidence that the magnetosphere couples to the disk outside
the corotation radius, possibly involving the so-called ``dead disk'' 
region \citep{dangelo2010, dangelo2012}.  We note, however, that 
$r_{\rm in}<r_{\rm co}$ is still allowed by the uncertainty in the $r_{\rm in}$
measurement.

\subsection{Accretion rate, luminosity, and source distance}

We can use our measured disk radius during steady pulsar spin-up to infer the
mass accretion rate $\dot M$ and hence the distance to the source. For steady 
long-term spin-up via magnetic accretion torques, we can assume that the 
magnetospheric radius $r_m$ lies well inside $r_{\rm co}$, so that the spin derivative 
of the pulsar should obey
\begin{equation}
2\pi\,I\,\dot\nu = \dot M \sqrt{G\,M_x\,r_m}\, ,
\end{equation}
where $\dot\nu$ is the pulsar's spin frequency derivative and $I$ is the pulsar's
moment of inertia.  Based on monitoring with
the {\em Fermi}/GBM 
instrument\footnote{See \url{http://gammaray.nsstc.nasa.gov/gbm/science/pulsars/lightcurves/4u1626.html}} 
\citep{camero2010}, the pulsar's spin-up rate during our 2010 observation was $\dot\nu = 4.4\times 10^{-13}$~Hz~s$^{-1}$.  
If we include the uncertainty arising from the correlation with $i_d$, our measured inner disk radius from the 2010 observation was $r_{\rm in}=(3.7^{+1.8}_{-0.7})\times 10^8\,M_{1.4}$~cm. Taking $r_m = r_{\rm in}$, our 2010 line fit implies a mass accretion rate of
\begin{eqnarray}
\dot M & = & (1.1^{+0.1}_{-0.2})\times 10^{16}\, I_{45}\, M_{1.4}^{-1}
                \mbox{\rm\ g~s$^{-1}$} \nonumber \\
  & = & (1.7^{+0.2}_{-0.4})\times 10^{-10}\, I_{45}\, M_{1.4}^{-1}\  M_\odot\mbox{\rm\,yr$^{-1}$} ,
\end{eqnarray}
and an X-ray luminosity of
\begin{equation}
L_x = \frac{G M_x \dot M}{R} = 
    (2.0^{+0.2}_{-0.4})\times 10^{36}\, I_{45} R_{10}^{-1}
    \mbox{\rm\,erg s$^{-1}$},
\end{equation}
where $I_{45}$ is $I$ in units of $10^{45}$~g~cm$^2$ and $R=10\, R_{10}$\,km is the neutron star radius.  We can compare this to the unabsorbed 0.1--10~keV X-ray flux  
$F_x = 5.3\times 10^{-10}$~erg~cm$^{-2}$~s$^{-1}$ measured in our 2010 observation. Based on a 2015 {\em NuSTAR} observation in the same spectral state \citep{dai2017}, we estimate the bolometric correction factor to be $f_{\rm bol}=2.5\pm 0.1$. We can then
determine the source distance to be 
\begin{equation}
D = (3.5^{+0.2}_{-0.3})\, I_{45}^{1/2} R_{10}^{-1/2} \left(\frac{f_{\rm bol}}{2.5}\right)^{-1/2} \mbox{\rm\ kpc}.
\end{equation}
Because $D$ depdends on the magnetospheric radius only as $r_m^{-1/4}$, we are able to derive a relatively precise distance despite the significant uncertainty on $r_{\rm in}$.  Given the source's Galactic latitude of $b=-13^\circ$, a 3.5~kpc distance means that 4U~1626$-$67 lies 0.8 kpc out of the Galactic plane. Our distance agrees with the optical parallactic distance of $3.5^{+2.3}_{-1.3}$~kpc from {\em Gaia} \citep{bailerjones2018}. It does not lie within the 5--13~kpc distance range inferred from optical reprocessing of the X-ray flux \citep{chakrabarty1998}, but that appears to be at least partially because of the assumption of $\cos i\approx 1$ made in that study.  If we repeat the \citet{chakrabarty1998} optical reprocessing analysis with our X-ray--fitted value of $\cos i=0.78$, we obtain a marginally consistent distance range of 3.5--10~kpc.

\subsection{Ionization conditions\label{ioncond}}

The presence of strong collisional ionization in the inner disk is surprising, given that it is illuminated by X-ray emission from the accreting neutron star. From simple energetics, we expect radiative heating to
dominate internal viscous heating at the accretion disk photosphere
for disk radii beyond
\begin{equation}
r  \gtrsim 1.6\times 10^8\,M_{1.4} 
                \left(\frac{1 - \eta}{0.1}\right)^{-1}
                \left(\frac{\sin\theta}{0.1}\right)^{-1} 
                \left(\frac{\varepsilon_x}{0.2}\right)^{-1} 
                \mbox{\rm cm} \,,
\end{equation}
where $\eta$ is the X-ray albedo, $\theta$ is the grazing angle of the
incident illumination of the disk by the neutron star, and
$\varepsilon_x$ is the accretion efficiency \citep{chakrabarty1998,jimenez2002}. This suggests that photoionization should be important throughout the accretion disk in 4U~1626$-$67.

It is instructive to make a simple calculation of the ionization conditions of the plasma. In steady-state, each ion $Z^{+i}$ with atomic number $Z$ and charge $i$ obeys the ionization balance equation,
\begin{multline}
    n_{Z,i}[\zeta_{Z,i} + n_e(\alpha_{Z,i}+\alpha^D_{Z,i}+C_{Z,i})] = \\
     n_{Z,i+1}\,n_e(\alpha_{Z,i+1}+\alpha^D_{Z,i+1}) \\
     + n_{Z,i-1}(\zeta_{Z,i-1}+n_e\,C_{Z,i-1}) , 
\label{eq:rates}
\end{multline}
where $\zeta$ is the photoionization rate, $\alpha(T_e)$ is the radiative recombination 
rate coefficient, $\alpha^D(T_e)$ is the dielectronic recombination rate coefficient, 
$C(T_e)$ is the collisional ionization rate coefficient, $n_{Z,i}$ is the number density of ion $Z^{+i}$, $n_e$ is the electron number density, and $T_e$ is the electron temperature. We neglect three-body recombination, which can become important at high densities \citep[$n_e>10^{17}$~cm$^{-3}$;][]{Bautista1998}. The terms on the left-hand side of equation~(\ref{eq:rates}) describe transitions out of state $i$,
and the terms on the right-hand side describe transitions into state $i$ from states
$i+1$ and $i-1$, respectively. The photoionization rate in a plasma at a distance $r$ from
an illuminating X-ray point source is
\begin{equation}
\zeta_{Z,i} = \frac{4\pi D^2}{4\pi r^2} \int_{E_{Z,i}^{\rm ion}}^\infty \left(\frac{dN}{dE}\right)_0\,\sigma_{Z,i}(E)\, dE ,
\end{equation}
where $\sigma(E)$ is the photoionization cross-section, $E^{\rm ion}$ is the ionization energy, $(dN/dE)_0$ is the unabsorbed illuminating photon continuum spectrum that we measure, and $D$ is our distance from the source. 

\begin{figure}
\includegraphics[angle=0,width=0.47\textwidth]{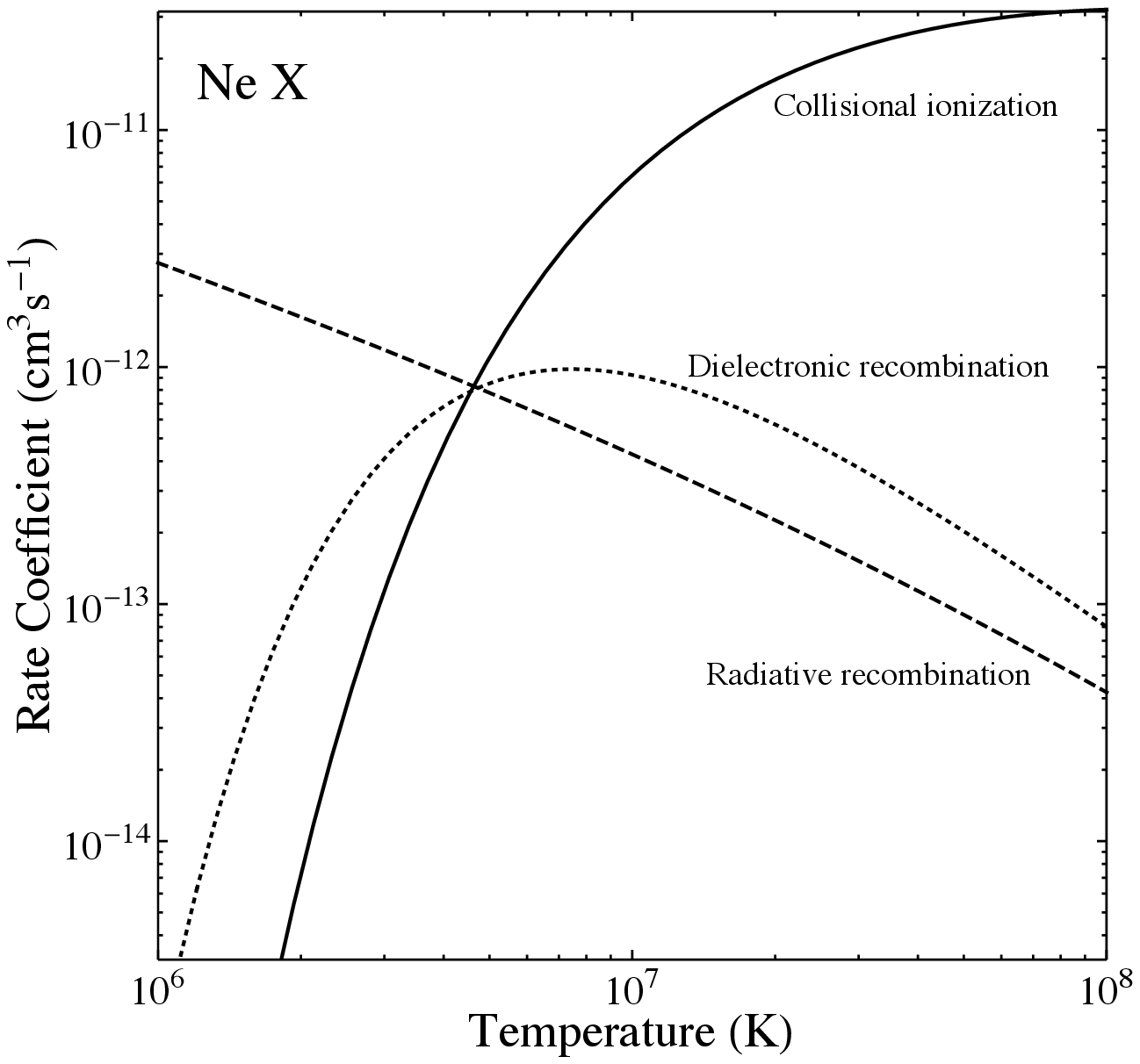}
\caption{\ion{Ne}{10} rate coefficients for collisional ionization \citep{ArnaudRothenflug1985,Mazzotta1998}, radiative recombination \citep{VernerFerland1996}, and dielectronic recombination \citep{Mazzotta1998}, as a function of temperature.
\label{RateFig}}
\end{figure}

\begin{deluxetable*}{lccccl}[t]
\tablecaption{Ionization, Recombination, and Abundances for $T_e=10^7$~K \label{tab:rates}}
\tablehead{Parameter & \colhead{Units} & \colhead{\ion{Ne}{9}} & 
  \colhead{\ion{Ne}{10}} & \colhead{\ion{Ne}{11}} & Ref.
  }
\startdata
Photoionization rate\tablenotemark{a}, $\zeta_{z,i}$ & s$^{-1}$ & $2.3\times 10^6$ & $0.9\times 10^6$ & $\cdots$ & 1 \\
Recombination rate coeff., $\alpha_{z,i}(T_e)$ & cm$^3$~s$^{-1}$ & $1.8\times 10^{-13}$ & 
     $4.3\times 10^{-13}$ & $1.2\times 10^{-12}$ & 2 \\
Dielectronic recombination rate coeff., $\alpha^D_{z,i}(T_e)$ & cm$^3$~s$^{-1}$ & $1.3\times 10^{-12}$ &
     $9.2\times 10^{-13}$ & $\cdots$ & 3 \\
Collisional ionization rate coeff., $C_{z,i}(T_e)$ & cm$^{3}$~s$^{-1}$ & $1.9\times 10^{-11}$ & 
     $6.4\times 10^{-12}$ & $\cdots$ & 4, 3 \\
Abundance fraction\tablenotemark{b} relative to \ion{Ne}{10}, $n_i/n_{\rm Ne\,X}$& $\cdots$ & 0.07 & 1.0 & 9.0 & 3 \\
Ionization energy, $E_{\rm ion}$ & eV & 1196 & 1362 & $\cdots$ & 1 \\
\enddata
\tablerefs{(1) \citet{VernerYakovlev1995}; (2) \citet{VernerFerland1996}; (3) \citet{Mazzotta1998}; (4) \citet{ArnaudRothenflug1985}.}
\tablenotetext{a}{Computed using $r=3.7\times 10^8$~cm, $D$=3.5~kpc, and $(dN/dE)_0$ for the 2010 observation from Table~\ref{tab:contspec}.}
\tablenotetext{b}{Computed assuming collisional ionization equilibrium.}
\end{deluxetable*}

\begin{figure}
\includegraphics[angle=0,width=0.47\textwidth]{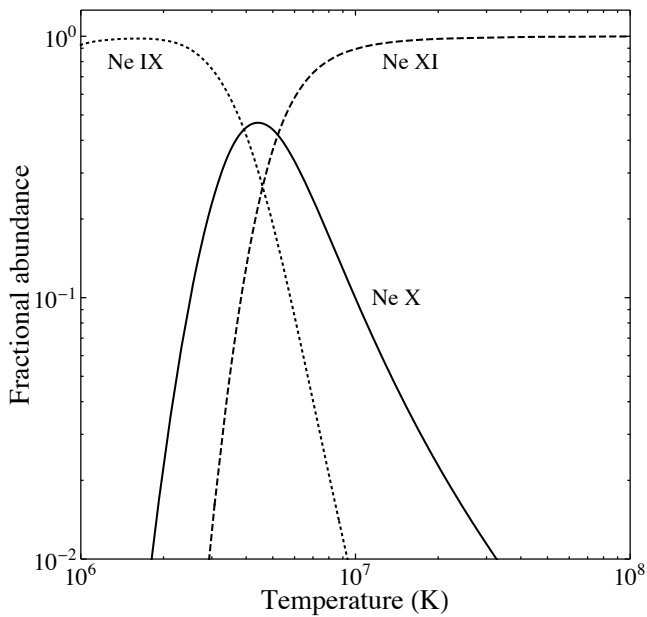}
\caption{Ionization balance for highly-ionized Ne as a function of temperature in collisional ionization equilibrium \citep{Mazzotta1998}.
\label{NeIonAbundFig}}
\end{figure}

\begin{figure}
\includegraphics[angle=0,width=0.47\textwidth]{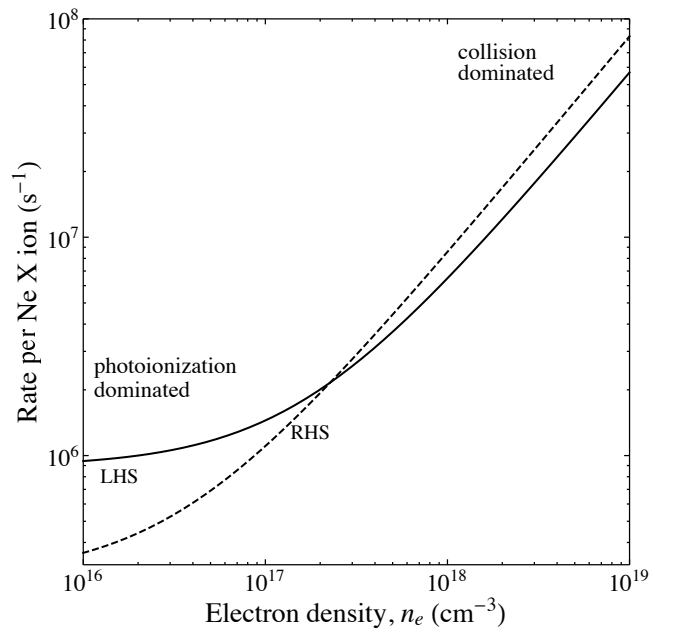}
\caption{The left-hand (solid curve) and right-hand (dashed curve) sides of equation~(\ref{eq:rates}) in the case of our 2010 observation of \ion{Ne}{10}.  The curves are dominated by photoionization at $n_e$ and by collisional ionization at high $n_e$. The slight displacement of the two curves at high $n_e$ is due to the approximations in our calculation (see text).  The two curves should be essentially identical in the high-$n_e$ limit, where the solution is independent of $n_e$ and valid as long as the plasma remains optically thin. The intersection shown thus reflects a rough lower limit on $n_e$ For our geometry, we calculate in \S\ref{sec:EM} that the plasma is optically thin for $n_e<6\times 10^{18}$~cm$^{-3}$. 
\label{EquationFig}}
\end{figure}

Let us consider the case of hydrogen-like \ion{Ne}{10} during the bright 2010 spin-up observation. If we assume $T_e=10^7$~K and take the observed 2010 continuum spectrum from Table~\ref{tab:contspec} with distances $r=3.7\times 10^8$~cm and $D=3.5$~kpc, then we obtain the rate coefficients in Table~\ref{tab:rates}. An examination of these coefficients reveals two important points. First, from the temperature dependence of the rate coefficients (see Figure~\ref{RateFig}), we see that collisional ionization is only competitive with recombination for $T_e\gtrsim 3\times 10^6$~K.  Second, given the high photoionization rate ($\sim 10^6$~s$^{-1}$), collisional ionization can only play a significant role for high densities $n_e\gtrsim 10^{16}$~cm$^{-3}$.  Note that if photoionization is neglected ($\zeta_{Z,i}=\zeta_{Z,i-1}=0$), then equation~(\ref{eq:rates}) is independent of $n_e$, which divides out. In that case, the ionization balance is set by $T_e$ via the rate cofficients, and the resulting equilibrium is valid for any $n_e$, as long as the plasma remains optically thin.

In our case, however, we know that photoionization must be taking place (from the irradiation), but we also know that collisional ionization is dominant and that $T_e\approx 10^7$~K (from our spectral fit). In order to use equation~(\ref{eq:rates}) to determine $n_e$, we first need to specify the relative abundances of the different ionization states of Ne ($n_{Z,i+1}/n_{Z,i}$ and $n_{Z,i-1}/n_{Z,i}$).  For simplicity, we assume the ionization balance corresponding to collisional ionization equilibrium, shown in Figure~\ref{NeIonAbundFig} \citep{Mazzotta1998}. Figure~\ref{EquationFig} then plots the two sides of equation~(\ref{eq:rates}). The flat portion of the curves at low density correspond to the regime where photoionization dominates.   The steep portion of the curves at high density correspond to the regime where collisional ionization dominates.  Strictly speaking, the two curves should lie on top of each other in this regime. They are slightly displaced from one another in our figure because our calculation is not completely self-consistent: our ionization balance from \citet{Mazzotta1998} neglects photoionization, and the rate coefficients they used to compute those relative abundances were not necessarily identical to the ones we used in Table~\ref{tab:rates}.  

In actuality, the solution to equation~(\ref{eq:rates}) does not correspond to a single intersection of the two curves in Figure~\ref{EquationFig}.  Rather, the equation admits a range of solutions with the two curves joining into a single curve extending to high density, where photoionization is negligible and the equation becomes independent of $n_e$.  (At sufficiently high density, the plasma will no longer be optically thin and our model breaks down.) Within the uncertainties, the intersection we see in Figure~\ref{EquationFig} thus corresponds to an approximate lower limit on the electron density, $n_e\gtrsim 2\times 10^{17}$~cm$^{-3}$. At this density, the photoionization rate is comparable to the collisional ionization rate, but with an ionization parameter of only
$\log \xi\approx 1.4$. We obtain similar results for \ion{O}{8}. This temperature and density combination is consistent with the atmosphere of an X-ray heated neutron-star accretion disk in an LMXB at a radius of order $10^8$~cm \citep{jimenez2002}.  If we repeat the same calculation for the faint spin-down state, using an average of the 2000 and 2003 continuum spectra and taking $T_e=8\times 10^6$~K and $r=8.3\times 10^8$~cm, we find $n_e\gtrsim 1\times 10^{16}$~cm$^{-3}$.  The lower density is to be expected, given the smaller ionizing luminosity and the lower electron temperature.

\subsection{Emission measure\label{sec:EM}}

We can calculate the emission measure of the hot component of the
line-emitting region by summing over the normalizations of the center and both wings of the lines in Table~\ref{tab:coll}.  For the 2010 spin-up observation,
\begin{equation}
 \mbox{\rm EM} = \cos i_d\,\int \beta\,n_e^2\,dV = 
   3.2\times 10^{57}\, D_{3.5}^2 \mbox{\rm\ \ cm$^{-3}$} ,
   \label{EM1}
\end{equation}
where $D_{3.5}$ is the source distance in units of 3.5~kpc, and we have used
equation~(\ref{eq:EMbeta}) and also accounted for the projection effect of the disk inclination $i_d$.  Similarly, for the 2003 spin-down observation,
\begin{equation}
 \mbox{\rm EM} = \cos i_d\,\int \beta\,n_e^2\,dV = 
   3.4\times 10^{56}\, D_{3.5}^2 \mbox{\rm\ \ cm$^{-3}$} .
   \label{EM2}
\end{equation}
We assume that the emitting region is an optically thin annulus of inner radius $r_{\rm in}$, width $\Delta r$,
thickness $\Delta h$, and volume $V\sim 2\pi\,r_{\rm in}\,\Delta h\,\Delta r$ in the disk 
atmosphere, on the visible side of the disk.  The radius $r_{\rm in}$ is determined by our 
{\tt diskline} fits, yielding $r_{\rm in}/r_{\rm co}$ of 0.5 and 1.3 for the spin-up and spin-down 
observations, respectively. We can use the same fits to estimate $\Delta r$ from the line 
emissivity index $q$ by assuming
\begin{equation}
 \left(\frac{r_{\rm in}+\Delta r}{r_{\rm in}}\right)^q = \frac{1}{e} ,
\end{equation}
yielding $\Delta r/r_{\rm in} \approx 0.3$. Finally, we assume $\Delta h/r_{\rm in} \approx 0.01$ from modeling of X-ray heated accretion disk atmospheres \citep{jimenez2002}.  This gives a volume of 
\begin{equation}
V  =  6.6\times 10^{23} \left(\frac{r_{\rm in}/r_{\rm co}}{0.5}\right)^3
                        \left(\frac{\Delta r/r_{\rm in}}{0.3}\right)
                        \left(\frac{\Delta h/r_{\rm in}}{0.01}\right)
        \mbox{\rm cm$^3$}
\label{V1}
\end{equation}
for the 2010 spin-up observation and 
\begin{equation}
V  =  1.2\times 10^{25} \left(\frac{r_{\rm in}/r_{\rm co}}{1.3}\right)^3
                        \left(\frac{\Delta r/r_{\rm in}}{0.3}\right)
                        \left(\frac{\Delta h/r_{\rm in}}{0.01}\right)
        \mbox{\rm cm$^3$}
\label{V2}
\end{equation}
for the 2003 spin-down observation.  

For the spin-up case, combining equations~(\ref{EM1}) and (\ref{V1}) gives
\begin{eqnarray}
\beta\, &\approx& \,0.6\ D_{3.5}^2\,n_{17}^{-2} 
                        \left(\frac{\cos i_d}{0.8}\right)^{-1}
                        \left(\frac{r_{\rm in}/r_{\rm co}}{0.5}\right)^{-3} \nonumber\\
 & & \quad \times \left(\frac{\Delta r/r_{\rm in}}{0.3}\right)^{-1}
                        \left(\frac{\Delta h/r_{\rm in}}{0.01}\right)^{-1} ,
\label{eq:EM_beta}
\end{eqnarray}
where $n_{17}$ is $n_e$ in units of $10^{17}$~cm$^{-3}$.  As we show in Appendix~B, a value of $\beta<0.83$ is unphysical, so at least one of the fiducial parameter values in equation~(\ref{eq:EM_beta}) requires adjustment. We estimated a rough lower limit of $n_{17}\gtrsim 2$ from the ionization conditions in \S\ref{ioncond}, although this is probably uncertain by a factor of a few, owing to the approximations we made. The requirement that the effective optical depth $\tau_{\rm eff}=\sqrt{\tau_{\rm es}\tau_{\rm ff}}$ for the collisional plasma is less than unity yields an upper limit
\begin{equation}
 n_{17} < 60\,\bar{g}_{\rm ff}^{-1/3} \left(\frac{\Delta h/r_{\rm in}}{0.01}\right)^{-2/3} ,
\end{equation}
where $\tau_{\rm es}$ is the optical depth for electron scattering, $\tau_{\rm ff}$ is the optical depth for free-free absorption (computed for the wavelength of the \ion{Ne}{10} line), and $\bar{g}_{\rm ff}$ is the (order-unity) velocity-averaged Gaunt factor \citep{rybicki1979}.  The formal requirement that $\beta>0.83$ also yields an upper limit of $n_{17}<0.9$, assuming the volume calculation above. Taken together, our measurements favor values of roughly unity for both $n_{17}$ and $\beta$.

\subsection{Nature of the mass donor}

Previous studies have shown that there are three types of Roche-lobe--filling mass donors possible for a 42~min binary period: (1) a 0.02$M_\odot$ degenerate dwarf with binary inclination $i\lesssim 33^\circ$; (2) a $0.08 M_\odot$ partially-degenerate, H-depleted star with $i\lesssim 8^\circ$; and (3) a $0.6 M_\odot$ He-burning star with $i\lesssim 1.3^\circ$ \citep{levine1988, verbunt1990,chakrabarty1998}. Given the inclination constraints, the {\em a priori} probabilities of these three possibilities are 16\%, 1\%, and 0.03\%, respectively. If we assume that the inner disk inclination and the binary inclination are 
identical ($i = i_d$), then we can use our {\tt diskline} fits from \S\ref{linespec} to discriminate between these possibilities. From equation~(6) of \citet{chakrabarty1998}, the stringent X-ray timing limits on the pulsar's orbit imply
\begin{equation}
\sin i < 7.8\times 10^{-3}\,q^{-1}(1+q)^{2/3}\,M_{1.4}^{-1/3}\,P_{42}^{-2/3},
\end{equation}
where $M_x = 1.4\,M_{1.4}\,M_\odot$ is the neutron star mass, $M_c$ is the companion
mass, $q=M_c/M_x$ is the binary mass ratio, and $P_{\rm orb} = 42\,P_{42}$~min is the
binary period.  Our spectral-fit value of $\sin i_d=0.63^{+0.23}_{-0.15}$ thus yields $q\lesssim
0.013$, consistent with a $\simeq$0.02~$M_\odot$ degenerate donor. \citet{chakrabarty1998} noted that such a donor would need to lie at a distance of $\lesssim$1~kpc if its mass transfer rate were driven by angular momentum loss due to gravitational radiation,
\begin{equation}
\dot M_{\rm gw} = 1.2\times 10^{-11}\,M_{1.4}^{8/3}\,
        \left(\frac{q}{0.01}\right)^2\ M_\odot\mbox{\rm\,yr$^{-1}$} ,
\end{equation}
the usual assumption for ultracompact binaries \citep[see review by][]{verbunt1995}. However, our measured mass accretion rate is an order of magnitude larger than $\dot M_{\rm gw}$.  The higher rate might be driven by X-ray heating of the mass donor \citep{lu2017}. Whatever the reason, this eliminates any discrepancy with our distance measurement. 

Our emission measure analysis favors a $\beta$ value of around unity. From Appendix~B, this is consistent with a highly-evolved, H-poor main sequence remnant or a He white dwarf donor. A C-O white dwarf donor would have $\beta\gtrsim 40$, and an O-Ne white dwarf donor would have $\beta\gtrsim 100$. \citet{heinke2013} have also argued for a He dwarf donor on the basis of binary evolution and disk stability arguments. The only evidence against this is the absence of detectable H and He lines in optical spectroscopy \citep{werner2006,nelemans2006}.  However, it is possible that the lines are weak or absent due to most or all of the H and He being completely ionized. Deeper optical spectroscopic measurements should be able to place stronger constraints on the H and He content of the donor.

\subsection{Energetics of the line emission}

The luminosity of the collisional plasma in the inner disk atmosphere (including both the line emission and the associated bremsstrahlung continuum) is
\begin{equation}
L_{\rm coll} = 3.4\times 10^{34} D_{3.5}^2\,\mbox{\rm erg~s$^{-1}$} .
\end{equation}
The associated cooling time $t_c = (3/2)n_e kT_e V/L_{\rm coll}$ is of order milliseconds, so a continuous energy source is required.  The gravitational energy of the accretion flow at the inner disk edge,
\begin{equation}
 \dot E_{\rm acc} \sim \frac{G M_x \dot M}{r_{\rm in}} = 
            5\times 10^{33}\,M_{1.4}^{-1}
            \left(\frac{r_{\rm in}/r_{\rm co}}{0.5}\right)^{-1}
            \mbox{\rm\ erg~s$^{-1}$} ,
\end{equation}
is an order of magnitude too small.  Magnetic reconnection in the disk truncation region is another possible energy source.  Assuming that the pulsar magnetic field is dipolar,
\begin{equation}
 B(r) = 3\times 10^{12} \,R_{10}^3\,
      \left(\frac{r}{\mbox{\rm 10 km}}\right)^{-3}
      \mbox{\rm\ G} ,
\end{equation}
we can estimate the rate of energy extraction through magnetic reconnection as
\begin{eqnarray}
 \dot E_{\rm mag} & = & \frac{B^2(r_{\rm in})}{8\pi} v_{\rm A}\,
        4\pi r_{\rm in}\, \Delta r_{\rm sheet} \\
   & = & 9\times 10^{31} R_{10}^6 \left(\frac{v_{\rm A}}{c}\right)
            \left(\frac{r_{\rm in}/r_{\rm co}}{0.5}\right)^{-4} \nonumber \\
   & & \quad \times \left(\frac{\Delta r_{\rm sheet}/r_{\rm in}}{10^{-5}}\right)
        \mbox{\rm\ erg s$^{-1}$}
\end{eqnarray}
where $v_{\rm A}$ is the Alfven velocity and $\Delta r_{\rm sheet}$ is the width of the current sheet \citep{degouveia2010}. This is over two orders of magnitude too small.  However, we note that $L_{\rm coll}$ is comparable to the fraction $f$ of the total pulsar luminosity $L_x$ intercepted at normal incidence by the optically thick inner edge of the accretion disk,
\begin{eqnarray}
 f L_x & = & \frac{(2\pi r_{\rm in})(2 h_{\rm thick})}{4\pi r_{\rm in}^2} L_x \\
  & = & 2\times 10^{34} \left(\frac{h_{\rm thick}/r_{\rm in}}{0.01}\right)
       \mbox{\rm\,erg~s$^{-1}$} ,
\end{eqnarray}
where $h_{\rm thick}$ is the height of the optically-thick disk. This suggests that direct X-ray heating of the inner disk edge may be responsible for powering the line emission, with the emitting region lying in the atmosphere just above the disk edge. 

This may partially explain why 4U~1626$-$67 is the only NS/LMXB in which strong collisional emission lines are observed from the inner accretion disk.  The H-poor composition of the donor does not seem to be relevant, as a H-rich solar-abundance plasma with the same density and temperature parameters would be a strong line source as well. Instead, we suggest that it is a combination of having a strongly magnetized pulsar, a low $L_x$, and a low binary inclination that leads to observable Keplerian disk lines.  Most NS/LMXBs contain accretion disks that extend all the way into their weakly magnetized neutron stars. The disk surface is heated by grazing-incidence X-rays, and neither the energetics nor the ionization conditions that we observed in 4U~1626$-$67 are reproduced. A magnetically truncated disk may be necessary in order to achieve the required direct radiative heating of the disk mid-plane. If $L_x$ is too high, then photoionization might completely ionize the inner disk atmosphere, eliminating line emission. (This could also occur if the pulsar's magnetic field strength is too weak, leading to a smaller inner disk radius.)  And finally, a low inclination angle \cite[$\lesssim 60^\circ$;][]{frank1987} allows a direct line of sight to the inner disk region. This orientation has an {\em a priori} probability of 50\% for an ensemble of binaries with isotropically-distributed orbital angular momentum vectors. 

There are few other strongly magnetized pulsars among LMXBs.  The 1.24~s accreting pulsar Her X-1 should experience similar direct heating of its truncated disk edge. However, the eclipsing nature of this source establishes it as a high-inclination binary \citep[$i\gtrsim 80^\circ$;][]{joss1984}, so the atmosphere of its inner accretion disk is presumably blocked from our direct view by the flaring of its outer disk. Instead, the photoionization spectrum observed from the Her~X-1 arises in the outer disk and a disk corona \citep{jimenez2005,ji2009}. Similarly, the 0.59~s accreting pulsar 4U~1822$-$371 is also observed at high inclination \citep[$i\simeq 82^\circ$;][]{hellier1989, heinz2001} and has a complex photoionization spectrum that arises in the disk corona and the impact point of the accretion stream onto the disk \citep{cottam2001,ji2011}. 

The 0.467-s bursting pulsar transient GRO J1744-28 is a more promising candidate, as it may have a small inclination \citep[$i\lesssim 10^\circ$;][]{finger1996,daumerie1996}. However, its inclination may be too low to detect a Keplerian profile. We note that an {\em XMM-Newton}/EPIC-pn CCD observation of the source near the peak of its 2014 outburst ($L_x=2\times 10^{38}$~erg~s$^{-1}$) found evidence for a complex of broadened, highly ionized emission lines of Si, Ar, Ca, and Fe that were ascribed to photoionization and possibly disk reflection \citep{dai2015}.  A higher-resolution HETGS observation later in the outburst ($L_x=8\times 10^{37}$~erg~s$^{-1}$) detected only a broad Fe fluorescence line \citep{degenaar2014}. It would be interesting to consider whether any of the emission features from the earlier observation might instead arise in a collisionally ionized atmosphere near the inner disk edge, although the high $L_x$ makes the conditions very different than in 4U~1626$-$67.  An observation much later in an outburst, when $L_x\sim 10^{36}$~erg~s$^{-1}$, would be most likely to find conditions similar to 4U~1626$-$67.

\acknowledgments
We thank all the members of the {\em Chandra} team for their enormous efforts,
and especially D. P. Huenemoerder, J. Davis, and J. Houck for their help with
HETGS data processing and fitting procedures.  D.C. also thanks Adam
Foster, Randall Smith, and Patrick Slane for help in understanding the APEC spectral models; Dimitrios Psaltis and Paul Hemphill for useful discussions; and Avi Loeb for hosting his sabbatical stay at the Harvard-Smithsonian Center for Astrophysics.  Support for this work was provided in part by the National 
Aeronautics and Space Administration (NASA) through Smithsonian Astrophysical
Observatory (SAO) contract SV3-73016 to MIT for support of the {\em Chandra}
X-ray Center and {\em Chandra} science instruments. CXC is operated by SAO
for and on behalf of NASA under contract NAS8-03060.

\facility{CXO (HETGS)}

\software{{\tt CIAO} \citep{fruscione2006}, 
        {\tt ISIS} \citep{houck2000},
        {\tt XSPEC} \citep{arnaud1996},
        {\tt XSTAR} \citep{kallman2001},
        {\tt APED} \citep{smith2001}}

\appendix
\section{Ultraviolet De-excitation of He-Like Triplets}

In the He-like triplet lines of \ion{O}{7} and \ion{Ne}{9} observed in
4U~1626$-$67, the intercombination line ($i$) is much stronger than the
forbidden line ($f$). Normally, a small value for the line ratio
$f/i$ indicates a very high electron density. However, another
possibility is that the upper level of the forbidden transition is
depopulated by a strong ultraviolet (UV) continuum into the upper
levels of the intercombination lines \citep{porquet2010}. As it
happens, there is a strong UV continuum (from the accretion disk) present in
4U~1626$-$67 \citep{homer2002}. Here, we demonstrate that UV de-excitation
prevents the use of the $f/i$ ratio as an accurate density diagnostic in this 
source.  Our calculation follows the analysis by \citet{marshall2013} for SS~433. 

The forbidden line in the He-like triplets arises from the $2\,^3S_1
\rightarrow 1\,^1S_0$ transition and has a spontaneous decay rate of \citep{smith2001}
\begin{equation}
A_f = \left\{
\begin{array}{ll}
9.77\times 10^3 \mbox{\rm\ s$^{-1}$} & \mbox{\rm for \ion{Ne}{9}} \\
9.12\times 10^2 \mbox{\rm\ s$^{-1}$}& \mbox{\rm for
  \ion{O}{7}}. \\ \end{array} \right. 
\end{equation}
These downward forbidden transitions will compete with UV photoabsorption in
upward $2\,^3S_1 \rightarrow 2\,^3P_J$ transitions, with total angular
momentum quantum number $J=\{0,1,2\}$. The photoabsorption rate for an
upward transition $1\rightarrow 2$ is 
\begin{equation}
R_{\rm pa} = B_{12} J_\nu = \frac{c^2}{2h\nu^3}\frac{g_2}{g_1}A_{21}
J_\nu ,
\end{equation}
where $A_{21}$ and $B_{12}$ are the Einstein coefficients; $g_1$ and $g_2$
are statistical weights for the lower and upper states, respectively; and $J_\nu$ is the mean intensity of the
incident radiation field \citep{rybicki1979}.  The relevant parameters for these
transitions are given in Table~\ref{tab:uv}. 

\begin{deluxetable}{cccccccc}[t]
\tablecaption{ATOMIC TRANSITION DATA FOR He-LIKE IONS \label{tab:uv}}
\tablewidth{0.47\textwidth}
\tablehead{ & & &\multicolumn{2}{c}{Ne IX} & & 
  \multicolumn{2}{c}{O VII}\\ \cline{4-5}\cline{7-8}
  \colhead{State} & \colhead{$g$} & & \colhead{$\lambda$ (\AA)} & 
  \colhead{$A_{21}$ (s$^{-1}$)} & &
  \colhead{$\lambda$ (\AA)} &   \colhead{$A_{21}$ (s$^{-1}$)} 
}
\startdata
2\,$^3S_1$ & 3 & &\nodata &\nodata & & \nodata & \nodata  \\ \\
2\,$^3P_0$ & 1 & &1277.7 & $1.01\times 10^8$ && 1639.9 & $7.97\times 10^7$ \\
2\,$^3P_1$ & 3 & &1272.8 & $1.05\times 10^8$ && 1638.3 & $8.1\times 10^7$ \\
2\,$^3P_2$ & 5 & &1248.1 & $1.13\times 10^8$ && 1623.7 & $8.41\times 10^7$ \\
\enddata
\tablecomments{All values are from the APED database \citep{smith2001}, \url{http://atomdb.org}}
\end{deluxetable}

The incident UV field arises from the accretion disk
surface. We model the disk as flat and thin, with inner radius
$r_1$ and outer radius $r_2$. (A flared disk would simply amplify the UV
illumination.) We assume that the emission line region
is located at a height $z$ above the inner disk annulus at $r=r_1$. Let us
consider a point in this region, located at cylindrical coordinates
$(r,\phi,z) = (r_1, 0, z)$.  Viewed from this point, the solid angle
subtended by an area element on the disk surface at $(r, \phi, 0)$ is
\begin{equation}
d\Omega = \frac{dA\,\cos\theta}{R^2} = \frac{zr\,dr\,d\phi}{R^3} ,
\end{equation}
where $\theta$ is the angle between the line of sight and the vertical
(parallel to the $z$-axis), and
$R=(r^2-2r_1r\cos\phi+r_1^2+z^2)^{1/2}$ is the line-of-sight distance
from the point to the area element. The mean intensity is then 
\begin{eqnarray}
J_\nu & = & \frac{1}{4\pi} \int I_\nu\,d\Omega \\
&=& \frac{1}{4\pi}\int_0^{2\pi}\int_{r_1}^{r_2}
   \frac{B_\nu(T[r])\,zr\,dr\,d\phi}{(r^2-2r_1r\cos\phi+r_1^2+z^2)^{3/2}} ,
\end{eqnarray}
where $I_\nu$ is the specific intensity, $B_\nu(T)$ is the Planck
function, 
\begin{equation}
 B_\nu = \frac{2h\nu^3/c^2}{e^{h\nu/kT}-1} ,
\end{equation}
and the temperature is 
\begin{equation}
T(r) = \left(\frac{3GM\dot M}{8\pi\sigma r^3}\right)^{1/4}
\end{equation}
for a viscously heated disk (neglecting X-ray heating\footnote{X-ray heating of the disk in 4U~1626$-$67 will begin to dominate viscous heating at $r\gtrsim 10^9$~cm \citep{chakrabarty1998}. This will tend to amplify the amount of UV deexcitation.}).  For nominal
values of $r_1=3.7\times 10^8$~cm, $r_2=2\times 10^{10}$~cm,
$M=1.4\,M_\odot$, $\dot M=1.7\times 10^{-10} \dot M_\odot$~yr$^{-1}$, and
$z=0.02\,r_1$ in 4U 1626$-$67 (see \S\ref{sec:EM}), we find through numerical integration that
\begin{equation}
J\nu = \left\{
\begin{array}{ll}
0.021 \mbox{\rm\ erg\,cm$^{-2}$\,s$^{-1}$\,Hz$^{-1}$\,sr$^{-1}$} &
\mbox{\rm at 1250\,\AA}\\ 
0.015 \mbox{\rm\ erg\,cm$^{-2}$\,s$^{-1}$\,Hz$^{-1}$\,sr$^{-1}$} &
\mbox{\rm at 1630\,\AA} \\ \end{array} \right.
\end{equation}
for the wavelengths corresponding to \ion{Ne}{9} and \ion{O}{7},
respectively.  In order to compute the total photoabsorption rate, we
must sum over the three possible upper states
\begin{eqnarray}
R_{\rm pa} & = & \sum_{J=0}^{2} B_{12,J}\,J_\nu(\nu_J) \\
 & = & \frac{c^2}{2g_1 h}
\sum_{J=0}^{2}\frac{g_{2,J}\,A_{21,J}}{\nu_J^3}\,J_\nu(\nu_J). 
\end{eqnarray}
where the atomic transition data are given in Table~\ref{tab:uv}. 
Finally, we can compute the ratio of the photoabsorption rate to the
forbidden line decay rate,
\begin{equation}
\frac{R_{\rm pa}}{A_f} = \left\{
\begin{array}{ll}
3500 & \mbox{\rm for \ion{Ne}{9}} \\
4.4\times 10^4 & \mbox{\rm for \ion{O}{7}} .\\ \end{array} \right.
\end{equation}
As noted above, we neglected flaring of the accretion disk height with radius as well as X-ray heating of the outer disk. Both of these effects would only tend to increase $R_{\rm pa}$. Given that $R_{\rm pa}/A_f \gg 1$, we find that UV photoabsorption dominates for both ions, rendering the He-like triplets unusable as a density diagnostic in 4U~1626$-$67.  Their use as a {\em temperature} diagnostic, however, is unaffected. 

\bigskip

\section{Emission Measure in a Hydrogen-Depleted Collisional Plasma}

In this Appendix, we show how to correctly interpret the fit
normalization from the APEC spectral model for optically-thin,
collisionally ionized plasmas \citep{smith2001} for the case of
a plasma containing no H or He. 

\subsection{Background}

For simplicity, we consider a two-level system (with upper
level $k$ and lower level $j$) of an element with atomic number $Z$ in
ionization state $p$ in a plasma.  The volume line emissivity
(in units of ph~cm$^{-3}$~s$^{-1}$) of the downward radiative
transition is 
\begin{equation}
  P_{kj} = n_k\,A_{kj} ,
\label{eq:Pkj}
\end{equation}
where $n_k$ is the number density in energy level $k$ and $A_{kj}$ is the
Einstein $A$ coefficient for spontaneous emission in the $k\rightarrow
j$ transition (in units of s$^{-1}$). The corresponding line flux (in
units of ph~cm$^{-2}$~s$^{-1}$) for an extended source region at
distance $D$ is 
\begin{equation}
  F_{kj} = \frac{\int P_{kj}\,dV}{4\pi D^2} .
\end{equation}
We can express $n_k$ in terms of the plasma conditions \citep{mewe1999},
\begin{equation}
  n_k = \left(\frac{n_k}{n_{Z,p}}\right)
        \left(\frac{n_{Z,p}}{n_Z}\right)
        \left(\frac{n_Z}{n_{\rm H}}\right)
        \left(\frac{n_{\rm H}}{n_e}\right) n_e ,
\end{equation}
where $n_{Z,p}$, $n_Z$, $n_{\rm H}$, and $n_e$ are the number densities
for ionization state $p$, element $Z$, hydrogen, and electrons,
respectively. On the right hand side, the first factor represents the
level population, the second factor the ionization balance, the
third factor the elemental abundance, and the fourth factor the
hydrogen-to-electron ratio.  Neglecting radiative excitations and 
stimulated emission, the collisional equilibrium condition at
temperature $T$ is 
\begin{equation}
  n_j\,n_e\,q_{jk}(T) = n_k\,A_{kj} + n_k\,n_e\,q_{kj}(T),
\end{equation}
where $q_{jk}(T)$ and $q_{kj}(T)$ are the collisional rate
coefficients (in units of cm$^{3}$~s$^{-1}$) for electron-impact
excitation and de-excitation, respectively.  Except at very high
density\footnote{For the \ion{Ne}{10} Ly$\alpha$ line at 12.13~\AA\ at
  $T=10^7$~K, the conditions $n_e\, q_{kj} \ll A_{kj}$ and 
  $A_{kj} \gg n_e\, q_{jk}$ are both satisfied if $n_e \ll 10^{23}$~cm$^{-3}$.
  We can generally assume that any optically-thin collisional plasma obeys
  $n_e\ll n_{\rm crit}$, where $n_{\rm crit}= A_{kj}/q_{kj}$ is the so-called
  critical density.}, 
the de-excitation term on the right-hand side is negligible
and $A_{kj} \gg n_e\,q_{jk}$, so that nearly all the ions are in the
ground state $j$ (that is, $n_k \ll n_j$).  We can hence write
\begin{eqnarray}
  n_{Z,p} & = & n_j + n_k \\
          &\approx & n_j .
\end{eqnarray}
We can then re-write equation~(\ref{eq:Pkj}) as
\begin{eqnarray}
  P_{kj} & = & n_j\,n_e\,q_{jk} \\
         & = & n_{Z,p}\,n_e\,q_{jk} .
\end{eqnarray}

\subsection{Ordinary hydrogen-rich plasmas}

Ordinarily, hydrogen is by far the dominant ion species in an
astrophysical plasma.  It is then convenient to write the volume line emissivity as
\begin{equation}
 P_{kj} = n_e\,n_{\rm H}\,
              \left[ \left(\frac{n_{Z,p}}{n_Z}\right)
                     \left(\frac{n_Z}{n_{\rm H}}\right)_\odot q_{jk}\right]
                     \mbox{\rm Ab}(Z) ,
\label{eq:Pkj2}
\end{equation}
where 
\begin{equation}
  \mbox{\rm Ab($Z$)} = \left(\frac{n_Z}{n_{\rm H}}\right)\bigg/ \left(\frac{n_Z}{n_{\rm H}}\right)_\odot
\end{equation}
is the number abundance of element $Z$ relative to H, expressed as a fraction of the solar abundance ratio $(n_Z/n_{\rm
  H})_\odot$.  The factor in square brackets in equation~(\ref{eq:Pkj2}) is the line emissivity
$\varepsilon_{kj}(T)$ per unit $n_e$ and $n_{\rm H}$ (in units of
ph~cm$^{3}$~s$^{-1}$) in a solar-abundance plasma. (The APEC models use the APED atomic physics database\footnote{See \url{http://atomdb.org} and \cite{smith2001}.} to calculate level populations and $q_{jk}$, and hence $\varepsilon_{kj}(T)$, as a function of temperature.) 

In addition to line emission, a collisional plasma also emits bremsstrahlung (free-free) continuum radiation with volume emissivity (in units of erg~cm$^{-3}$~s$^{-1}$~Hz$^{-1}$)
\begin{equation}
 \varepsilon_\nu^{\rm ff} = 
   C_{\rm ff}(T)\,n_e\,n_{\rm H}\, \sum_Z C_Z(T)
        \left(\frac{n_Z}{n_{\rm H}}\right)_\odot\,\mbox{\rm Ab}(Z) ,
\label{eq:ff1}
\end{equation}
with
\begin{equation}
 C_{\rm ff}(T) = \frac{32\pi e^6}{3m_e c^3}\left(\frac{2\pi}{3m_e kT}\right)^{1/2}
        \,\exp(-h\nu/kT) ,
\end{equation}
and
\begin{equation}
 C_Z(T) = \sum_p Q_{Z,p}^2 \left(\frac{n_{Z,p}}{n_Z}\right) \bar{g}_{\rm ff}(Q_{Z,p},T),
\end{equation}
where $\nu$ is the emission frequency; $Q_{Z,p}$ is the effective charge of ionization state $p$ for element $Z$; $e$ and $m_e$ are the charge and mass of the electron; and $\bar{g}_{\rm ff}(Q_{Z,p},T)$ is the velocity-averaged Gaunt factor \citep{rybicki1979}. Even though we are discussing a H-rich plasma, we have explicitly written Equation~(\ref{eq:ff1}) to include the contribution of all the ions for completeness. Note that although the Gaunt factor is of order unity, it is a nonlinear function of $Q_{Z,p}^2$, so it can vary significantly for different ions and ionization states \citep{vanHoof2014}. This can be an important effect when H does not dominate the composition. We have assumed that the electrons and all the ions are at the same temperature. 

The fit parameters of the APEC models are temperature $T$ and the elemental abundance multipliers Ab($Z$). The fit normalization is proportional to the volume emission measure 
\begin{equation}
\mbox{\rm EM} = \int n_e\,n_{\rm H}\,dV .
\label{eq:EM}
\end{equation}
For a known volume and electron-to-hydrogen ratio, EM can be used to
determine $n_e$.  In an ordinary astrophysical plasma, nearly all the
electrons come from the ionization of H and He, with only a negligible
contribution from heavier elements.  For a solar-abundance plasma that
is hot enough for both H and He to be completely ionized,  
\begin{equation}
  n_e \simeq n_{\rm H} + 2 n_{\rm He} \simeq 1.2 n_{\rm H} .
\label{eq:solar}
\end{equation}
Metal abundances do not change the ratio $n_e/n_{\rm H}$ in a hydrogen-rich plasma by more than a few percent. 

\subsection{Hydrogen-depleted plasma}

In a hydrogen-depleted plasma, it is the heavier elements that provide
the unbound electrons.  A given emission measure will then imply a
different $n_e$ than in the hydrogen-rich case.  We can rewrite
the volume emissivities in terms of the overall ion density $n_i$ (which
includes all heavy elements), 
\begin{equation}
 n_i = \sum_Z n_Z .
 \end{equation}
For the volume line emissivity, equation~(\ref{eq:Pkj2}) becomes
\begin{equation}
 P_{kj} = n_e^2\, \left(\frac{n_i}{n_e}\right)
                          \left(\frac{n_{Z,p}}{n_Z}\right)
                          \left(\frac{n_Z}{n_i}\right)
                          q_{jk}  ,
\label{eq:Pkj3}
\end{equation}
while for the volume bremsstrahlung emissivity, equation~(\ref{eq:ff1}) becomes
\begin{equation}
 \varepsilon_\nu^{\rm ff} = C_{\rm ff}(T)\, n_e^2\, 
    \left(\frac{n_i}{n_e}\right)
    \sum_Z C_Z(T) \left(\frac{n_Z}{n_i}\right) .
\label{eq:ff2}
\end{equation}
We now re-write the emission measure in equation~(\ref{eq:EM}) as 
\begin{equation}
 \mbox{\rm EM} = \int \beta\,n_e^2\, dV .
\end{equation}
For a solar-abundance plasma, equation~(\ref{eq:solar}) yields $\beta =
0.83$. In cases where H and He are heavily depleted or absent,
however, $\beta$ must be computed using the fit values for the abundances and an assumed plasma composition and ionization state.  It is thus a correction factor specific to an observation, rather an intrinsic property of a particular plasma composition. We can compute $\beta$ from either the measured lines or the continuum. For the lines, comparing equations~(\ref{eq:Pkj2}) and (\ref{eq:Pkj3}) yields
\begin{equation}
  \beta_{\rm line} = \left(\frac{n_i}{n_e}\right)
          \left(\frac{n_Z}{n_i}\right)
          \left(\frac{n_Z}{n_{\rm H}}\right)_\odot^{-1}
          \frac{1}{\mbox{\rm Ab}(Z)} .
\label{eq:bline}
\end{equation}
For the continuum, comparing equations~(\ref{eq:ff1}) and (\ref{eq:ff2}) yields
\begin{equation}
  \beta_{\rm ff} = \left(\frac{n_i}{n_e}\right)
     \frac{\sum\limits_{Z} C_Z(T) \left(\cfrac{n_Z}{n_i}\right)}
          {\sum\limits_{Z} C_Z(T) \left(\cfrac{n_Z}{n_{\rm H}}\right)_\odot\mbox{\rm Ab}(Z)} .
\label{eq:bff}
\end{equation}
Since the same emission measure applies to the lines and the continuum, we expect
$\beta_{\rm line} = \beta_{\rm ff} \equiv \beta$. However, it is clearly more straightforward to compute $\beta$ from lines, since $\beta_{\rm ff}$ depends upon the detailed composition and ionization state of the entire plasma instead of just the ion-to-electron ratio of a single species.

\subsection{Application to 4U 1626$-$67}

Our observations of 4U 1626$-$67 detected lines of Ne and O in what is
presumed to be a plasma devoid of H. We will use equation~(\ref{eq:bline}) to compute 
$\beta$. From our collisional model fit parameters for the 2010 data, we can calculate 
that the relative abundance of Ne and O is
\begin{equation}
  \frac{n_{\rm Ne}}{n_{\rm O}} = 
     \frac{\mbox{\rm Ab(Ne)}}{\mbox{\rm Ab(O)}}\,
     \left(\frac{n_{\rm Ne}}{n_{\rm H}}\right)_\odot
     \left(\frac{n_{\rm O}}{n_{\rm H}}\right)_\odot^{-1} = 0.46\pm 0.14,
\end{equation}
where our measured values of Ab(Ne)=2.6$\pm$0.5 and Ab(O)=0.8$\pm$0.2 are taken from
Table~\ref{tab:coll}, and the solar abundance ratios are from
\citet{anders1989}. If we assume that the 
plasma consists of only Ne and O, then we can write
\begin{equation}
  n_i = n_{\rm O} + n_{\rm Ne} \simeq 1.46 n_{\rm O}.
\end{equation}
Approximating the plasma as fully ionized, we then have
\begin{equation}
  n_e = 8 n_{\rm O} + 10 n_{\rm Ne} = 12.6 n_{\rm O} ,
\end{equation}
and find that $n_i/n_e = 0.12$. Then, using either the Ne or O
fit abundances, equation~(\ref{eq:bline}) yields $\beta=120$. 

\begin{figure}[t]
\centerline{\includegraphics[angle=0,width=0.8\textwidth]{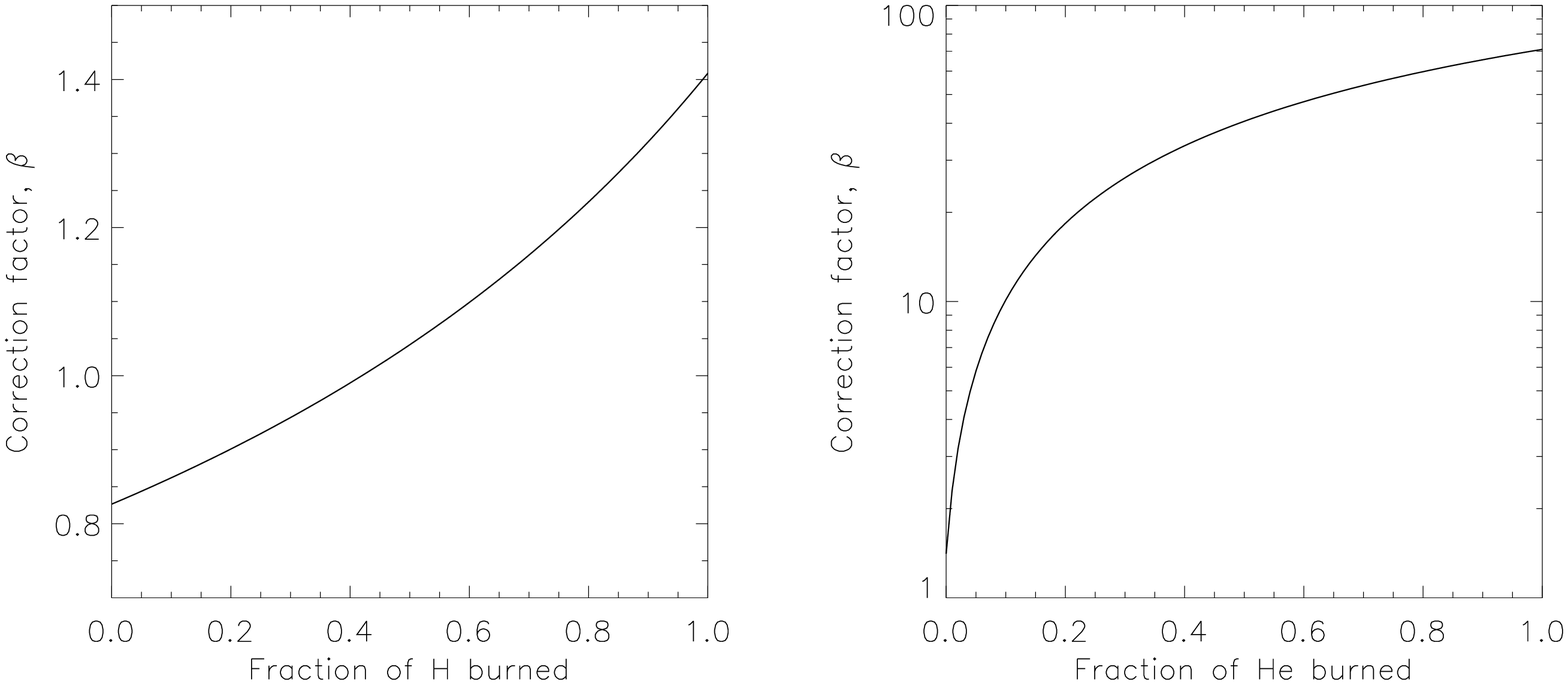}}
\caption{Emission-measure correction factor $\beta$ as a function of hydrogen (left) and helium (right) burned in a white dwarf progenitor, calculated for a Ne/O number ratio of 0.46.
\label{betafig}}
\end{figure}

We now consider the possibility that there is also C present.  Even though
there are no features of C in the HETGS bandpass, the presence of C would still affect the
electron population.  We take the mass fractions of C and O
to be equal \citep{segretain1994}.  This is equivalent to setting 
$12 n_{\rm C} = 16 n_{\rm O}$, or $n_{\rm C}/n_{\rm O} = 4/3$.  We then have
\begin{equation}
  n_i = n_{\rm C} + n_{\rm O} + n_{\rm Ne} 
  = n_{\rm O}(1.33+1+0.46)
  =  2.79 n_{\rm O} ,
\end{equation}
and for a fully-ionized plasma,
\begin{equation}
  n_e = 6 n_{\rm C} + 8 n_{\rm O} + 10 n_{\rm Ne} = 20.6 n_{\rm O} ,
\end{equation}
yielding $n_i/n_e = 0.13$. From equation~(\ref{eq:bline}), we find $\beta\simeq 70$.  

Finally, we consider the general case of an arbitrary donor composition. We start with a solar-composition progenitor and proceed through both hydrogen-burning and helium-burning phases while enforcing our observed Ne/O number ratio. We first calculate the evolution of $\beta$ as we burn the H to He via the (net) $pp$ reaction,
\[
  4\ \mbox{$^1$H} \rightarrow \mbox{$^4$He} + 2 e^+ + 2 \nu_e + 2\gamma .
  \]
After all the H is burned, we then calculate the evolution of $\beta$ as we burn He to C via the triple-$\alpha$ reaction,
\[
  3\ \mbox{$^4$He}  \rightarrow \mbox{$^{12}$C} + \gamma .
\]
During He burning, some of the resulting C will burn to O, and some of the O to Ne, via the reactions
\begin{eqnarray*}
  \mbox{$^{12}$C} + \mbox{$^4$He} & \rightarrow & \mbox{$^{16}$O} + \gamma \\
  \mbox{$^{16}$O} + \mbox{$^4$He} & \rightarrow & \mbox{$^{20}$Ne} + \gamma .
\end{eqnarray*}

We assume that the rate of C and O production is such that equal masses of each element are produced. Thus, for every 24 He atoms burned, we produce 4 C atoms and 3 O atoms. We further assume that the rate of Ne production is such that our observed Ne/O fraction of 0.46 by number is maintained. We approximate the plasma as fully ionized for the purposes of this calculation.  We then use equation~(\ref{eq:bline}) to calculate $\beta$.  The resulting evolution of $\beta$ and $n_i/n_e$ as a function of the fraction of H or He burned in the progenitor is shown in Figure~\ref{betafig}. In a H-burning progenitor, $\beta$ increases from 0.83 to 1.4. In a He-burning progenitor, $\beta$ further increases to around 70 by the time all the He is burned to C, O, and Ne.  

\bibliographystyle{aasjournal}
\bibliography{paper}

\end{document}